\documentclass[twocolumn]{aastex631}
\usepackage{graphicx}
\usepackage{epsfig}
\usepackage{natbib}
\usepackage{multirow}
\usepackage{amsmath}
\usepackage{textcomp}
\usepackage{ulem}
\shorttitle{Impact of Cosmic Variance on EoR Inference}
\shortauthors{Bruton et al. (2022)}

\newcommand{\g}{{\rm\thinspace g}}

\newcommand{\s}{{\rm\thinspace s}}

\newcommand{\muv}{$M_{\textrm{UV}}$}
\newcommand{\ewlya}{$\textrm{EW}_{\textrm{Ly}\alpha}$}

\newcommand{\lya}{Ly$\alpha$}

\newcommand{\G}{{\rm G}}
\newcommand{\K}{{\rm K}}
\newcommand{\hi}{H\thinspace{\sc i}}

\newcommand{\vpeaks}{$\Delta v_{\rm Ly\alpha}$}

\newcommand{\xf}{$\bar{\rm{x}}_{\rm{HI}}$}

\newcommand{\Msol}{\hbox{\thinspace M$_{\sun}$}}

\begin{document}

\title{The Impact of Cosmic Variance on Inferences of Global Neutral Fraction Derived from L\lowercase{y}$\alpha$ Luminosity Functions During Reionization}

\correspondingauthor{Sean Bruton}
\email{bruto012@umn.edu}

\author[0000-0002-6503-5218]{Sean Bruton}
\affiliation{Minnesota Institute for Astrophysics, University of Minnesota, 116 Church St SE, Minneapolis, MN 55455, USA}

\author[0000-0002-9136-8876]{Claudia Scarlata}
\affiliation{Minnesota Institute for Astrophysics, University of Minnesota, 116 Church St SE, Minneapolis, MN 55455, USA}

\author[0000-0003-3291-3704]{Francesco Haardt}
\affiliation{National Institute of Nuclear Physics INFN, Milano - Bicocca, Piazza della Scienza 3, 20126 Milano, Italy}
\affiliation{DiSAT, Università dell’Insubria, via Valleggio 11, 22100 Como, Italy}

\author[0000-0001-8587-218X]{Matthew J. Hayes}
\affiliation{Stockholm University, Department of Astronomy and Oskar Klein Centre for Cosmoparticle Physics, AlbaNova University Centre, SE-10691, Stockholm,
Sweden}

\author[0000-0002-3407-1785]{Charlotte Mason}
\affiliation{Cosmic Dawn Center (DAWN)}
\affiliation{Niels Bohr Institute, University of Copenhagen, Jagtvej 128, 2200 København N, Denmark}

\author[0000-0003-4965-0402]{Alexa M. Morales}
\affiliation{Department of Astronomy, The University of Texas at Austin, 2515 Speedway, Stop C1400, Austin, TX 78712, USA}

\author[0000-0003-3374-1772]{Andrei Mesinger}
\affiliation{Scuola Normale Superiore, Piazza dei Cavalieri 7, I-56126 Pisa, Italy}

\begin{abstract}
{We investigate the impact of field-to-field variation, deriving from cosmic variance, in measured Lyman-$\alpha$ emitter (LAE) luminosity functions (LFs) and this variation's impact on inferences of the neutral fraction of the intergalactic medium (IGM) during reionization. We post-process a z=7 IGM simulation to populate the dark matter halos with LAEs. These LAEs have realistic UV magnitudes, \lya\ fluxes, and \lya\ line profiles. We calculate the attenuation of \lya\ emission in universes with varying IGM neutral fraction, \xf. In a \xf$=0.3$ simulation, we perform 100 realizations of a mock 2 square degree survey with a redshift window $\Delta z = 0.5$ and flux limit $\rm{f}_{Ly\alpha}>1\times10^{-17}\:\rm{ergs}\:\: \rm{s}^{-1} \: \rm{cm}^{-2}$; such a survey is typical in depth and volume of the largest LAE surveys conducted today. For each realization, we compute the LAE LF and use it to recover the input \xf. Comparing the inferred values of \xf\ across the ensemble of the surveys, we find that cosmic variance, deriving from large-scale structure and variation in the neutral gas along the sightline, imposes a floor in the uncertainty of $\Delta \bar{\rm{x}}_{\rm{HI}} \sim 0.2$ when \xf $=0.3$. We explore mitigation strategies to decrease this uncertainty, such as increasing the volume, decreasing the flux limit, or probing the volume with many independent fields. Increasing the area and/or depth of the survey does not mitigate the uncertainty, but composing a survey with many independent fields is effective. This finding highlights the best strategy for LAE surveys aiming at constraining \xf\ of the universe during reionization.
}
\end{abstract}

\section{Introduction}\label{introduction}
\noindent
Reionization is the most recent phase change of the universe, in which the hydrogen in the intergalactic medium (IGM) went from being virtually fully neutral to nearly completely ionized. It is thought to have been complete within about the first billion years of the universe. It is expected to have been a patchy process, as ionized bubbles formed and grew around the first sources of ionizing radiation in the universe. Generally, it is thought that stars in the first galaxies provided the bulk of the necessary ionizing radiation to drive reionization \citep{finkelstein2015, robertson2015, bouwens2015a}, though which galaxies dominate the ionizing photon budget is still being investigated. 


Over the past several years, observations have been made which put constraints on the timeline of reionization using complementary methods sensitive to ionized and neutral hydrogen. Some examples: \citet{planck-collaboration2020} use the scattering signature of electrons freed during reionization on the cosmic microwave background to constrain the midpoint on instantaneous reionization to redshift $z_{re}=7.67\pm0.73$. The fraction of dark pixels in the \lya\ and $\textrm{Ly}\beta$ forests constrains reionization to be very nearly complete at $z\sim6$ \citep{becker2021,qin2021, bosman2022}. Quasar spectra at $z>7$ offer insight into the global neutral fraction ($\Bar{x}_{\textrm{HI}}$) of the IGM throughout reionization based on the IGM's \lya\ damping wing imprints \citep{greig2019,davies2018}.

Galaxies, and in particular \lya-emitters (LAEs) are also used to constrain the IGM hydrogen neutral fraction, and its evolution with time. \lya\ is a resonant transition in hydrogen and  \lya\ photons scatter multiple times even in small amounts of neutral gas \citep{dijkstra2014,dijkstra2017}. Thus, observations of \lya\ are particularly useful to probe the \hi\ spatial distribution, column density and velocity fields. Observations of  \lya\  freely propagating through the IGM is indicative of a largely ionized IGM or a significantly redshifted \lya\ line. 

The LAE fraction, the fraction of galaxies which are found to emit \lya, has been found to rapidly decline at $z>6$ \citep{stark2010, pentericci2011, jung2018}, and this decline is interpreted as a sign of increasing neutral hydrogen fraction in the IGM \citep{treu2013, mesinger2015, mason2018}. There are complications, however, in that the observed decreasing LAE fraction could also result from evolving galaxy properties, such as the escape fraction of \lya\ photons \citep{dijkstra2014}, an increase in the number of Lyman Limit Systems (LLS) \citep{bolton2013}, or other considerations (see \citet{finkelstein2016} for a review). Still, \citet{mesinger2015} and \citet{mason2018} argue that the increasing neutral fraction of the IGM is the most plausible explanation, and in this context the rapid disappearance of \lya\ at $z>6$ paints a picture of late and rapid reionization \citep{ouchi2018, hoag2019, mason2019, yoshioka2022}. Such a timeline for reionization may favor UV bright, massive galaxies being the drivers of reionization \citep{naidu2020}, as opposed to an undetected large population of faint objects \citep{finkelstein2019}. 

Many studies have also used the evolving luminosity function of LAEs to place constraints on reionization \citep{rhoads2001, malhotra2004, malhotra2006, hu2019, wold2021, morales2021}. However, the constraints on \xf\ at $z=7$ from LAE luminosity functions (LFs) are sometimes in slight tension. At $z=7$, \citet{zheng2017} find \xf $\sim0.40-0.60$, \citet{ota2017}\ find \xf $>0.4$, \citet{itoh2018} find \xf\ $=0.25^{+0.25}_{-0.25}$, \citet{hu2019} find \xf\ $\sim 0.2-0.4$, and \citet{wold2021} find \xf\ $<0.33$. More recently, \citet{morales2021} compared observed \lya\ LFs with models from an inhomogeneous reionization simulation to constrain \xf$\sim$ $0.08^{+0.08}_{-0.05}$, $0.28\pm0.05$, and $0.69\pm0.11$ at redshifts 6.6, 7.0, and 7.3, respectively. 

The moderate tension between these inferences on \xf\ can perhaps be attributed to cosmic variance. There is  stochasticity in the number of LAEs observed in a given survey arising from both large scale structure and the inhomogeneity of reionization.

It is noteworthy that LAEs at $z>6$ are fairly rare objects--dedicated narrowband surveys at $z\sim7$ typically detect tens of objects in volumes $\sim1\times10^6$ cMpc \citep{ota2017, itoh2018, hu2019, wold2021}; these small number statistics could easily lead to field-to-field variations in the measurement of the $z=7$ LAE LF. This effect has been observed in features like the LAE LF's ``bright end bump'' \citep{zheng2017, hu2019, wold2021}, the result of very bright LAEs falling within the field of observation. The ``bright end bump'' is not observed in all LAE surveys at $z=7$ (or even between different fields of the same survey), presumably either because of these objects inherent rarity or their concealment behind a neutral IGM. 

This is the crux of the issue we will explore in this paper: there is a degeneracy between the stochasticity in the number of LAEs observed and the inferred global neutral fraction of the universe. Indeed, \citet{mesinger2008b} showed that, owing to the inhomogeneity of reionization, there is intrinsic scatter in the inferred global neutral fraction of the universe from observations of the \lya\ damping wing in high-z quasars or gamma-ray bursts (GRBs). \citet{mcquinn2008a} similarly found that a single high-z GRB could not place a constraint on the global neutral fraction with an uncertainty better than $\Delta \bar{\rm{x}}_{\rm{HI}}\sim 0.3$. \citet{mason2018a} demonstrated the same principle with regards to \lya\ equivalent widths; inference on the global neutral fraction is inherently stochastic when using observations of the equivalent widths (EWs) of \lya\ in UV bright galaxies. Further, \citet{mesinger2008b},  \citet{mcquinn2008a}, and \citet{mason2018a} all showed that the uncertainty on \xf\ is a function of \xf\ itself, $\Delta \rm{\bar{x}_{HI}(\bar{x}_{HI})}$. Generally, the uncertainty tends to decrease as \xf\ increases because the size distribution of ionized bubbles gets smaller, since they have not yet had time to grow and merge in early reionization \citep{mason2018a}. We will demonstrate that there is a similar effect, a floor in the uncertainty on inferring \xf, when using the evolving Lya LF and explore mitigation strategies to reduce this uncertainty.




We post-process a cosmological inhomogeneous reionization simulation \citep{mesinger2007, mesinger2011, mesinger2016}, populating the dark matter halos with simulated galaxies from empirical relations. We first verify our simulation against the intrinsic UV and \lya\ luminosity functions (LFs) at $z\sim6$, then model the \lya\ LFs after IGM attenuation at $z\sim7$. We perform mock observations on our simulation, reproducing the probed volumes of surveys which are carried out today. Across each of the many realizations of these mock observational programs, we measure the LAE LF and infer \xf. We investigate how the inferred $z=7$ value of \xf\ changes as a result of statistical variance in the observed LAE LFs. 

In \S\ref{s2} we give an overview of the simulation and the post processing we have performed. In \S\ref{cosmic_variance} we explore the inherent uncertainty on inferred global neutral fraction at $z\sim7$ deriving from cosmic variance. We also explore if this uncertainty decreases by significantly increasing the survey area, depth, or using many independent fields. 
We conclude in \S\ref{conclusion}. Throughout the paper we use the AB magnitude system and the \citet{planck-collaboration2016} cosmology with ($\Omega_\Lambda$, $\Omega_m$, $H_0$) = (0.69, 0.31, 68 $km\:s^{-1} Mpc^{-1}$),

\section{The Reionization Simulation and its Post-Processing}\label{s2}
\noindent
In this section, we first summarize the simulation of dark matter halo masses, positions, and their \lya\ optical depths. We then present an overview of our methods for assigning galaxy properties to the halos.

\subsection{The 21cmFAST Simulation}
\noindent
We use custom simulations produced with the 21cmFASTv2 software \citep{mesinger2007, mesinger2011, mesinger2016}. 21cmFASTv2 calculates the evolution of the hydrogen neutral fraction in the early universe using a semi-numerical approach. Inside the simulation cube, 1.6 cGpc on each side with a $1024^3$ cell resolution, the code tracks dark matter and the phase of hydrogen in each cell while accounting for recombinations, photoheating star-formation suppression, supernova feedback, and radiation. Dark matter halos are identified from the $3072^3$ initial conditions, and mapped to Eulerian positions at z=7 using perturbation theory \citet{mesinger2007}. More information can be found in \citet{mesinger2016}. 

Each simulation contains the comoving Cartesian coordinate positions of halos with masses ranging from $10^{10.25} - 10^{12} \Msol$. The halos start with only 14 discrete halo masses, evenly spread in log space. To avoid affects resulting from this discretized mass function, we redistribute the masses to produce a smooth distribution in log space, preserving each halos positions while approximating the original mass. The simulation also computes \lya\ optical depths as a function of velocity offset from \lya\ line center, $\tau(\Delta v)$, up to $525\textrm{km}\:\textrm{s}^{-1}$ ($\Delta \lambda =2.1$ \AA) in steps of $75\textrm{km}\:\textrm{s}^{-1}$ ($\Delta \lambda \sim 0.3$ \AA). These optical depths are calculated by integrating the neutral hydrogen along a 300 comoving Mpc path.
Intrinsically, the optical depths are smooth functions of velocity offset from the line center, but we only have a coarsely sampled optical depths at particular velocity offsets. To mitigate any effect from this coarse sampling, we interpolate the optical depths across velocity offsets when calculating the IGM attenuation on the \lya\ line profile.

In total, we have nine simulations with different global neutral fractions, \xf, in the range $0.01 < \Bar{\rm{x}}_{\textrm{HI}} < 0.92$. Each simulation takes the same $z=7$ dark matter halo distribution and overlays on a different IGM ionization map, which corresponds to changing the ionization efficiency \citep{mcquinn2007a, mason2018}. To get an intuitive picture, one may imagine that in a scenario with large \xf\ (most IGM hydrogen neutral), any given halo will have a small ionized bubble around it, if any. In a low \xf\ scenario (most IGM hydrogen ionized), those same halos will have larger ionized bubbles around them, which may have begun to overlap with neighboring halos' bubbles. This means that each individual halo's optical depth, at a given velocity offset, smoothly increases as neutral fraction increases, which allows us to interpolate between the neutral fraction files, obtaining a finer grid in neutral fraction than our original nine simulations. 

\subsection{Propagating Halo Growth}\label{halo_growth}
\noindent
In order to make predictions at $z=7$, we  need to calibrate the \lya\ LF using observations at lower redshifts ($z\sim6$), where there is evidence that reionization is very nearly complete \citep{qin2021}. Though reionization may not be complete at $z\sim6$, residual \xf (i.e. if \xf$\sim0.1$) has a very small effect on the optical depth of \lya\ and so the LAE LF will not be significantly changed \citet{mason2018}. As redshift decreases, halos grow because they accrete matter through mergers. Thus, we need to forward propagate the halo mass distribution from $z=7$ to lower redshifts for proper comparison to observations.

We use the median halo growth trajectories from \citet{behroozi2019} to grow our halos. We interpolate between the trajectories to create a fine grid in halo mass, then assign each of the $z=7$ halo masses to a particular growth curve. To assign a growth trajectory to a halo mass, we find the curve which has $z=7$ halo mass closest to the given halo mass in the simulation. Then, moving along the matched growth curves to the lower redshift of interest, we get the new masses. These new masses are then redistributed into a smooth log-space distribution.

\subsection{Galaxy Properties}
\noindent
To achieve our goal of simulating LAEs, we need to assign intrinsic \lya\ luminosities and \lya\ line properties to the dark matter halos. Our road map will be: 
\begin{enumerate}
\itemsep0em 
  \item Assign absolute UV magnitudes (\muv) using a $M_h$-\muv\ relation (\S \ref{uv_relation})
  \item Assign a \lya\ EW (\ewlya) to each halo, using a UV dependent EW probability distribution function (PDF) (\S \ref{muv_ew})
  \item Assign \lya\ line profiles, with line velocity offset and width dependent on the halo masses (\S \ref{line_properties})
\end{enumerate}

Item 3 is required to compute the opacity to \lya\ photons from the neutral IGM. In what follows, we first describe the methodology used to assign \muv\ and \lya\ EW to each halo and how this methodology was validated with real data. We then discuss the new prescription to assign the \lya\ line profile and the comparison with observations. 

\subsubsection{The $M_h$-\muv\ Relation}\label{uv_relation}
\noindent
We use the $M_h$-\muv\ relation from \citet{mason2015} to assign UV magnitudes to our halos. This relation has an in-built redshift dependence, since halo distributions and their growth rates change with redshift \citep{mason2015}. 
We use relations tabulated at $z=6$ and $z=6.8$ in this paper. Considering our small range of redshifts of interest ($5.7 < z < 7$) and the relatively small evolution between the $z=6$ and $z=6.8$ $M_h$-\muv\ relation, we do not derive a new $M_h$-\muv\ relation for every redshift we consider. Instead, we apply the $z=6$ relation if our simulation redshift is $5.7<z<6.4$ and $z=6.8$ relation if our simulation redshift is $6.4<z<7$.

\begin{figure}[t!]
\epsscale{1.1}
\plotone{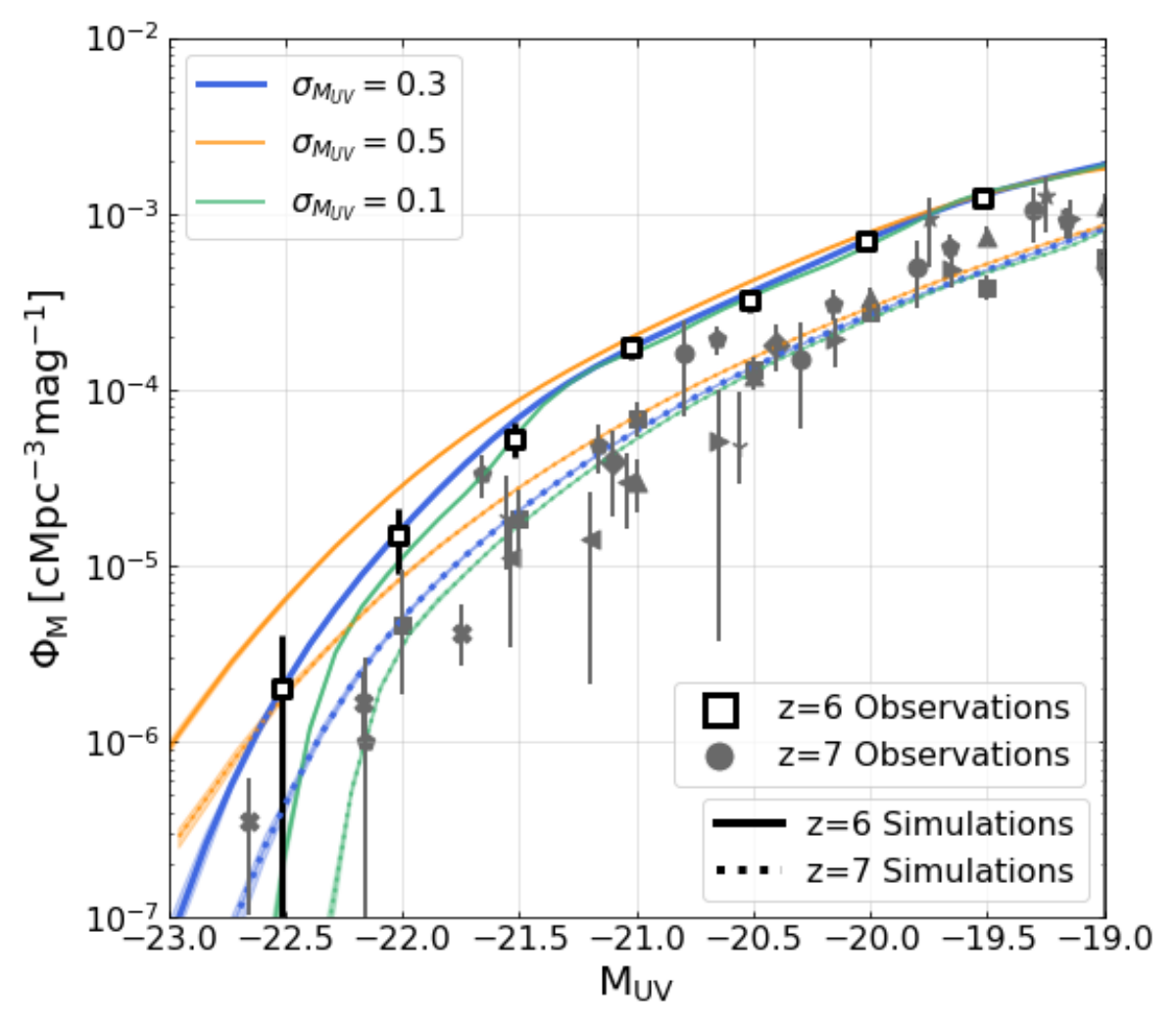}
\caption{Three $z=6$ UV LFs, resulting from three different $M_h$-\muv\ scatter values, are plotted in solid green ($\sigma_{M_{UV}} = 0.1$), blue ($\sigma_{M_{UV}} = 0.3$), and orange ($\sigma_{M_{UV}} = 0.5$). We calculate the UV LF for each $\sigma_{M_{UV}}$ value across 50 different realizations of the simulation and show the region within which 95\% of UV LFs fall with the shaded areas. The same color scheme is used for the z=7 UV LFs, which are plotted as dotted lines. The z=7 data are shown in gray; the squares are \citet{finkelstein2015}, stars are \citet{atek2015}, circles are \citet{bouwens2011}, pentagons are \citet{bouwens2015}, crosses are \citet{bowler2014}, diamonds are \citet{castellano2010}, down triangles are \citet{livermore2017}, up triangles are \citet{mclure2013}, left triangles are \citet{ouchi2009}, right triangles are \citet{schenker2013}, and tri-down are \citet{tilvi2013}. The $M_h$-\muv\ relation with scatter of $\sigma_{M_{UV}}=0.3$ reproduces the z=6 and z=7 data simultaneously. The bright end provides the demarcation between models with different $\sigma_{M_{UV}}$ values. \label{fig:uv_func_combined}}
\end{figure}

The relations provided in \citet{mason2015} do not include a prescription for the scatter, so we follow methods akin to \citet{ren2019} and \citet{whitler2020} and assume a normal scatter in \muv. The specific value of the scatter is chosen so that the resulting UV LF reproduces the measured LFs at $z\sim6$ and $z\sim7$ simultaneously. In Figure \ref{fig:uv_func_combined}, we show UV LF observations at z=6 in black and z=7 in gray. At both redshifts, we plot three UV LFs resulting from different scatter prescriptions: 0.1 in green, 0.3 in blue, 0.5 in orange. Solid lines are used for z=6, and dotted lines for z=7. We apply each relation with its scatter 50 times and indicate the range within which 95\%  of the UV LFs fall with the shaded regions.

Motivated by the agreement between the UV LFs with $\sigma_{M_{UV}} = 0.3$ and the observations at $z=6$ and $z=7$, we fix the value of $\sigma_{M_{UV}}$ to 0.3. The fact that we are able to reproduce UV LFs at two redshifts simultaneously with only one varying parameter is encouraging; it validates both our halo growth prescription and the $M_h$-\muv\ relation.

\subsubsection{\muv-\ewlya\ Distribution}\label{muv_ew}
\begin{figure}
\epsscale{1.1}
\plotone{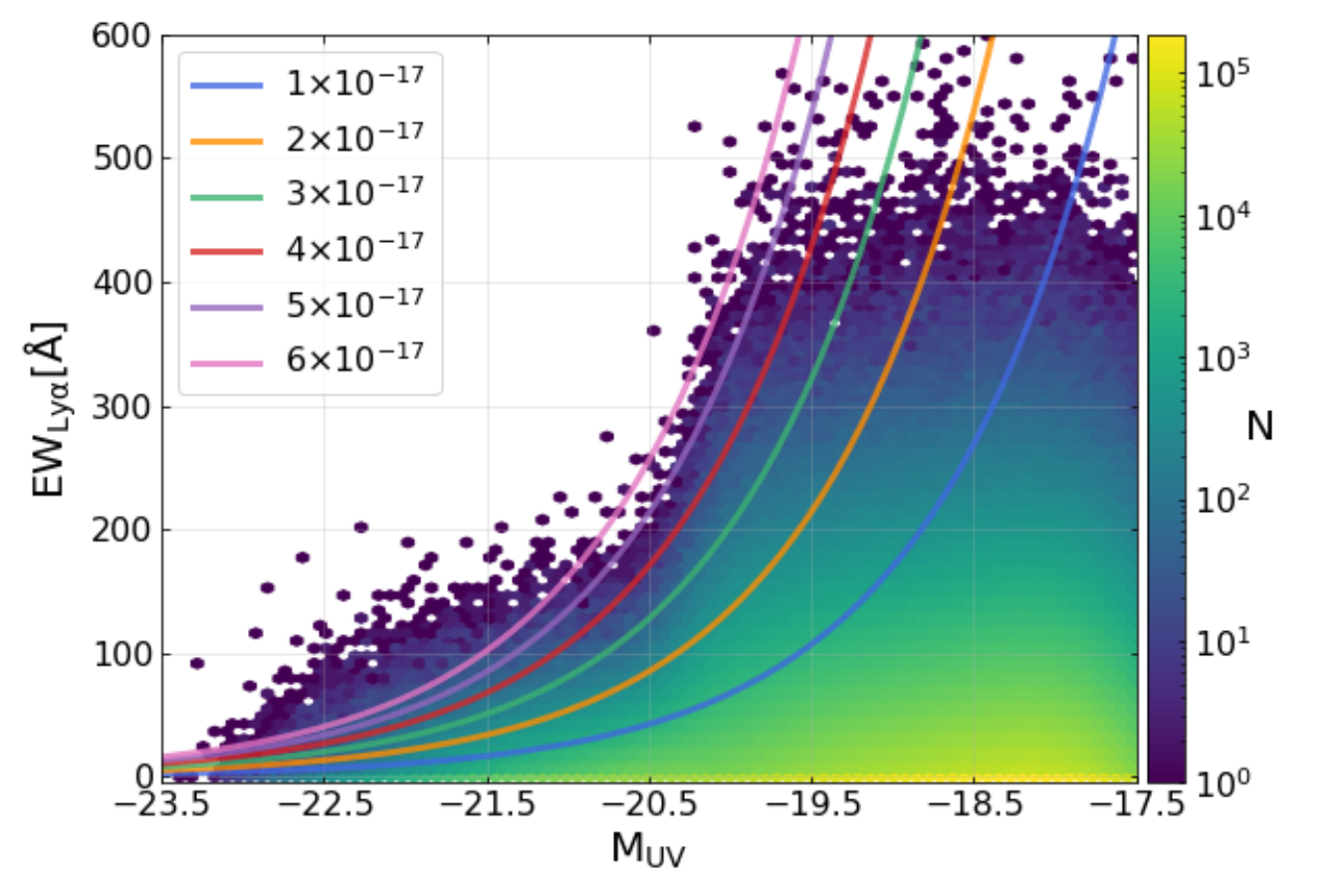}
\caption{The log 2D distribution of galaxies at z=6.6 in the $M_{UV}-\textrm{EW}_{\textrm{Ly}\alpha}$ plane. The colored lines denoted in the legend indicate traces of constant limiting \lya\ line flux in $\rm{ergs}\:\rm{s}^{-1}\:\rm{cm}^{-2}$. A survey with a given line flux limit could detect LAEs which lie above the trace corresponding to its flux limit. Note that there is an abundance of galaxies at \ewlya=0 along the very bottom of the distribution, corresponding to the non-\lya\ emitting population. \label{fig:mag_vs_ew}}
\end{figure}

\noindent
The distribution of rest-frame \ewlya\ given \muv\ takes the form
\begin{equation}
\begin{aligned}
p(\textrm{EW}_{\textrm{Ly}\alpha}\mid M_{\textrm{UV}}) = {} & \frac{A(M_{\textrm{UV}})}{W_c(M_{\textrm{UV}})}e^{-\frac{\textrm{EW}_{\textrm{Ly}\alpha}}{W_c(M_{\textrm{UV}})}}H(\textrm{EW}_{\textrm{Ly}\alpha})\\ & + [1-A(M_{\textrm{UV}})]\delta(\textrm{EW}_{\textrm{Ly}\alpha})
\end{aligned}
\end{equation}
from \citet{mason2018}. $H(\textrm{EW}_{\textrm{Ly}\alpha})$ is the Heaviside step function and $\delta({\textrm{EW}_{\textrm{Ly}\alpha}})$ is the Dirac delta function. \citet{mason2018} fit the parameters $A$ and $W_c$ using the \citet{debarros2017} measurements of \muv\ and \ewlya. The population from \citet{debarros2017} were photometrically selected using the Lyman break, so called Lyman break galaxies (LBGs), and it is worth noting that not all LBGs show \lya\ emission. The parameter $A$ accounts for the fraction of galaxies which do not emit \lya\ and $W_c$ determines the exponential decline of the probability distribution function towards larger \ewlya. Both parameters are allowed to vary as a function of \muv, and we take their value from \citet{mason2018}. To give a sense of the resulting distribution, the log 2D distribution of the \muv-\ewlya\ plane for $z=6.6$ galaxies, along with traces for constant \lya\ flux limits, are shown in Figure \ref{fig:mag_vs_ew}. UV faint galaxies have a larger probability of taking on a large \ewlya\ value than UV bright galaxies.

We are inherently assuming that the underlying \ewlya\ distributions are the same at $z\sim6$, where the \citet{debarros2017} observations are made, and at $z\sim7$, where we will apply the \ewlya\ distribution. \citet{mason2018} argue that, while oversimplifying, this assumption is justified by the relatively short time frame; less than 200 Myr pass between $z=6$ and $z=7$. Further, this means that any evolution in the \ewlya\  distribution between $z=6$ and $z=7$ would be attributed to an evolving \xf\ in our model, as explored in \citet{mason2018}.

\subsubsection{\lya\ Line Profile}\label{line_properties}
\noindent
Since the \lya\ optical depth due to the neutral IGM depends on the wavelength offset from the line center, we need to assign a \lya\ line profile that accounts for radiative transport effects in the galaxies' interstellar medium (ISM) and circumgalactic medium (CGM). We will take a relatively simple approach in modelling the \lya\ line profiles, modifying the methods in \citet{mason2018}. 

We assume that $z\sim6$ \lya\ line profiles are statistically the same as those at $z=7$ when ignoring attenuation from a neutral IGM. Under this assumption, we take observed \lya\ line properties at $z\sim6$, correlate them with observable galaxy properties, and assign them to our $z=7$ halos. The line profiles we will assign, then, have the effects of the ISM and CGM built-in. Attenuation by the IGM could change the observed \lya\ line profile statistics at $z=7$.

We assume that the line profile modified by the ISM and CGM is a Gaussian with an offset with respect to the rest-frame line center (\vpeaks). Unlike \citet{mason2018}, we assign the velocity offset with respect to the galaxy rest-frame using an empirical relationship from \citet{hayes2021}, which is explained further in the next subsection.

The line's full width half maximum (FWHM) is set equal to \vpeaks, as found in \citet{verhamme2018}. The \lya\ line profiles are truncated blueward of the \lya\ rest-frame line center because anything blueward of \lya\ line center will redshift into resonance and be scattered by residual neutral fraction. The \lya\ line profile after leaving the ISM and CGM is
\begin{equation}
     J_\alpha(\Delta v_{Ly\alpha}, M_h, v) \propto
    \begin{cases}
      \frac{1}{\sqrt{2\pi}\sigma_\alpha}e^{-\frac{(v-\Delta v_{Ly\alpha})^2}{2\sigma^2_\alpha}} & \text{if}\: v \geq 0\\
      0 & \text{otherwise}.\\
    \end{cases}    
\end{equation}

Note, though, that roughly one quarter of ultraluminous ($\rm{log}(\rm{L}_{Ly\alpha}/[ \rm{ergs}\:\rm{s^{-1}}]) > 43.5$) LAEs at $z>6$ emit flux blueward of line center \citep{hu2016, songaila2018, meyer2020}. These LAEs likely reside in large ionized bubbles in the IGM. These bubbles may have been carved out by ionizing radiation from the ultraluminous LAEs themselves, a population of nearby ionizing sources, an obscured quaser, or a combination of contributors (see \citet{bagley2017} and \citet{mason2020} for a discussion). Due to their rarity, we do not modify our line profiles to account for these ultraluminous galaxies.

\paragraph{\lya\ Velocity Offsets}\label{lya_offsets}
\noindent
\citet{hayes2021} compiled 229 \lya\ selected galaxies at $2.9 < z < 6.6$ from observations with the Very Large Telescope's (VLT) Multi Unit Spectroscopic Explorer (MUSE) and 74 galaxies at $z < 0.44$ from Hubble's Cosmic Origins Spectrograph (COS) and compared the two populations' \lya\ line profiles. The VLT galaxies' \lya\ velocity offsets relative to their systemic redshifts (i.e. the line profiles' 1st moments) were found by calibrating the observations with low-z data and a model \citep{runnholm2021}. These velocity offsets, \vpeaks, and the \lya\ luminosities are correlated. The data are described in more detail in \citet{hayes2021}. Note that these galaxies were selected with the \lya\ line, while the \citet{debarros2017} sample which provided the \ewlya\ distribution were selected with the Lyman break. While not all LBGs have \lya\ emission, we have captured this behavior in the \ewlya\ distribution; those that do have \lya\ emission may be detected in the \citet{hayes2021} sample, and so these empirical velocity offsets from \lya\ selected galaxies may be applied to LBGs as well. Note, however, that we find that the transmission fraction of \lya\ is largely insensitive to the exact velocity offset value in our model.

\begin{figure}
\epsscale{1.1}
\plotone{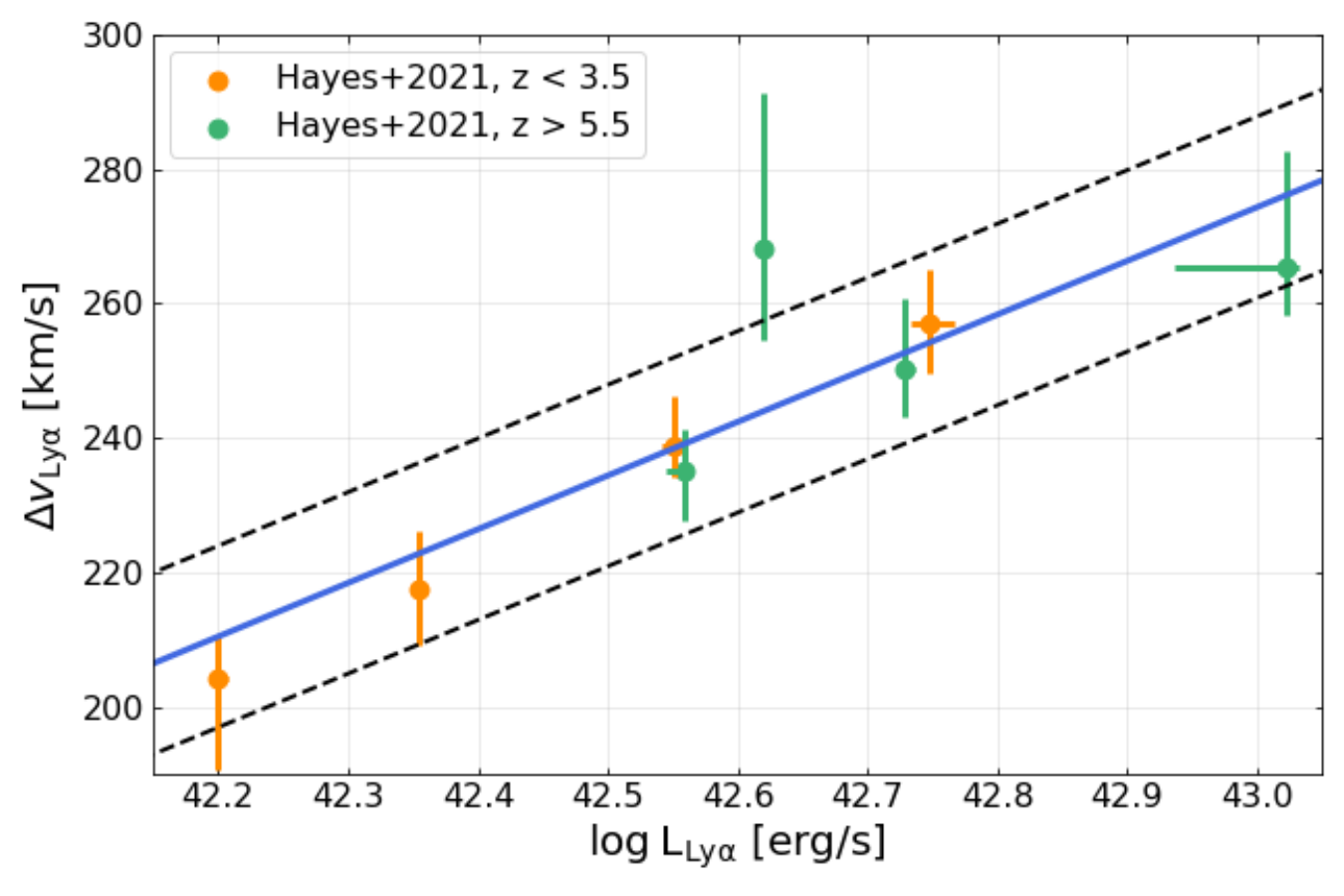}
\caption{We show the 1st moment of the \lya\ lines versus the log of the \lya\ luminosity for the MUSE data at $z<3.5$ (green) and $z>5.5$ (red). Our linear fit is shown with the blue line, while the black hashed lines show the $1-\sigma$ Gaussian scatter about the relation that we fit with \texttt{emcee}. \label{fig:1st_mom}}
\end{figure}

We use \texttt{emcee} to fit a linear relation to the observed high-z LAEs' line profiles' \vpeaks and the log of their \lya\ luminosities. The best fit linear relation is given by
\begin{equation}
    \Delta v_{\rm{Ly}\alpha} = -3156^{+882}_{-895} + 80^{+21}_{-21}\:\rm{log_{10}}(\rm{L}_{Ly\alpha})
\end{equation}
with a Gaussian scatter of $13.5^{+6.1}_{-3.5}$ km/s. The best fit relation, with the MUSE data, is shown in Figure \ref{fig:1st_mom}. The $z<3.5$ MUSE data are shown in orange, and the $z>5.5$ data are shown in green. Both redshift ranges are consistent with the same linear relation, indicating little to no redshift evolution. This linear relation is used to center the Gaussian before truncation; random Gaussian scatter of $13.5$ km/s is added to the offsets.

As we will only be concerned with \lya\ luminosities larger than $\rm{L}_{Ly\alpha} \gtrsim 10^{42} \:\rm{ergs}\:\rm{s}^{-1}$, the fact that our relation will predict a negative velocity offset at \lya\ luminosities $\lesssim 4\times 10^{39} \:\:\rm{ergs}\;\:\rm{s}^{-1}$ is not an issue.

With each halo assigned a \muv, \ewlya, \lya\ optical depth ($\tau_{\rm{IGM}}$), and \lya\ line profile, we have everything we need to compute the transmitted fraction of \lya\ flux. Explicitly, the fractional transmission through the IGM is:
\begin{equation}
    T = \frac{\int_0^{\infty}J_\alpha e^{-\tau_{\text{IGM}}}dv}{\int_0^{\infty}J_\alpha dv}.
\end{equation}

\subsection{Validation}
With all the pieces required to simulate \lya-emitters in place, we proceed to validate the recipes by comparing the simulation's $z=5.7$ \lya\ LF to the observations. At this redshift, the dark pixel fraction in the \lya\ and Ly$\beta$ forest indicates that reionization is nearly complete \citep{qin2021}, but even with a neutral fraction of \xf $\sim$ 10\% the impact on \lya\ transmission is negligible \citep{mason2018}. As such, this comparison will test only the $M_h$-\muv\  and  \muv-\ewlya\ relations.

\begin{figure}
\epsscale{1.1}
\plotone{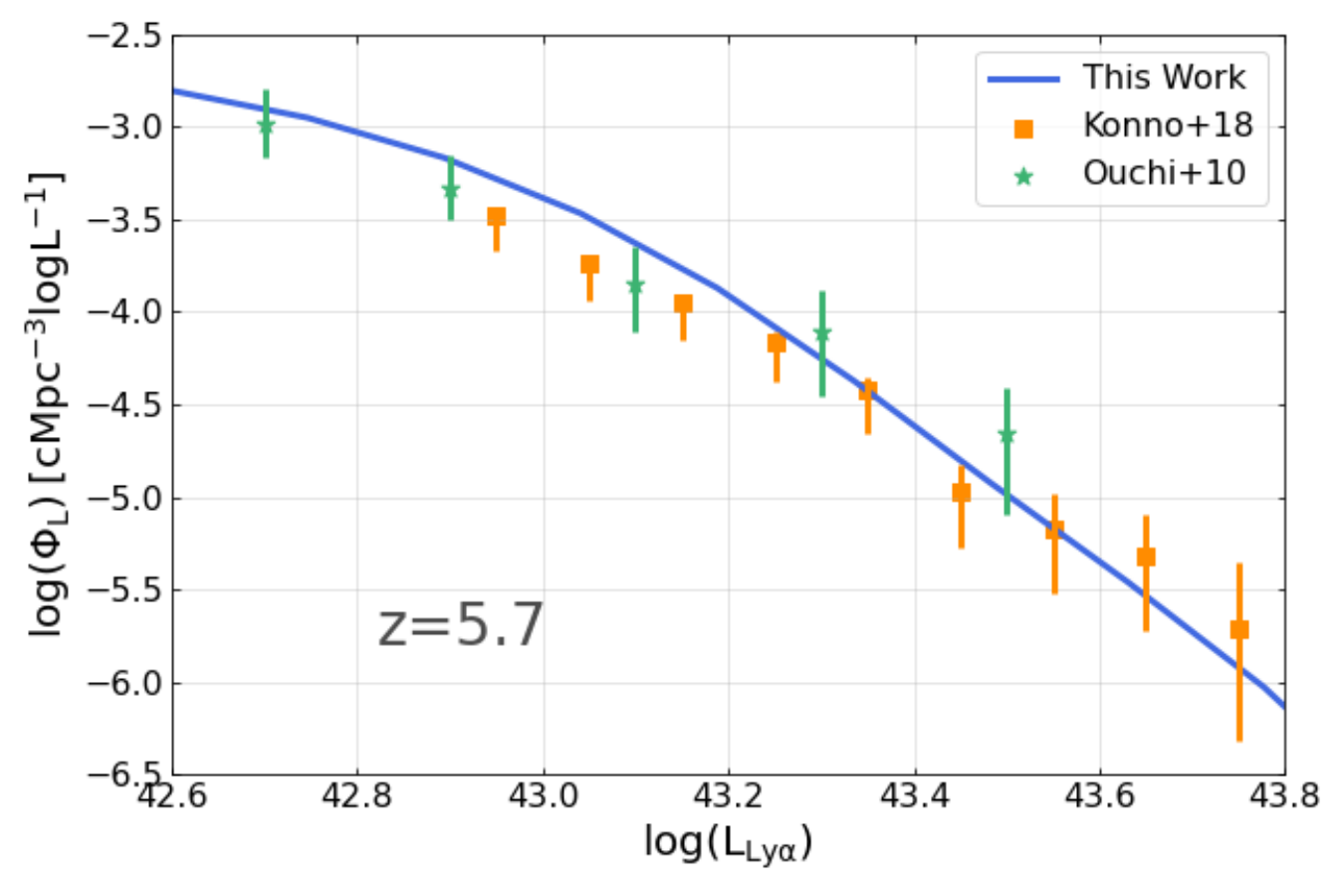}
\caption{The $z=5.7$ \lya\ LF is shown in blue with constraints from \citet{konno2018} and \citet{ouchi2010}. We see very good agreement between our simulation and the observational constraints. \label{fig:lya_lf_z57}}
\end{figure}

The $z=5.7$ \lya\ LF is shown in Figure \ref{fig:lya_lf_z57} along with observations from \citet{hu2010, ouchi2010, konno2018}. The halos have been grown to $z=5.7$ as explained in Section~\ref{halo_growth} and the UV luminosities are assigned from these evolved masses. The \ewlya\ are assigned via the relation described in Section \ref{muv_ew}. The resulting $z=5.7$ \lya\ LF is in very good agreement with observations, lending credit to the recipes we have used thus far.

This work will remain focused on the impact and mitigation of cosmic variance on the inference of \xf, particularly at $z=7$. A reader interested in analytic modelling of the \lya\ LFs at other redshifts should consult \citet{morales2021}, which uses similar methods as in this paper.

\subsection{Full Volume z=7 \lya\ LFs}\label{full_lfs}
We used the elements discussed in the previous sections to compute the $z=7$ \lya\ LFs from the full $\sim4.1\:\rm{cGpc}^3$ simulation volume, and for all values of \xf\ ranging from 0.01 to 0.92, with spacing of 0.03. Using the entire volume allows us to minimize shot noise in the range of \xf\ and \lya\ luminosity of interest. For each value of \xf, we recalculate the simulation 30 times. We take the median of the 30 realizations as the true model; this has the effect of further minimizing shot noise and sampling the sources of stochasticity from our empirical relations. The resulting \lya\ LFs are our best estimate of the true LAE LF at $z=7$ for various \xf\ values. In Section \ref{inference}, we will compare mock observations to these forward models to see how well the observations constrain \xf.

\begin{figure}
\epsscale{1.1}
\plotone{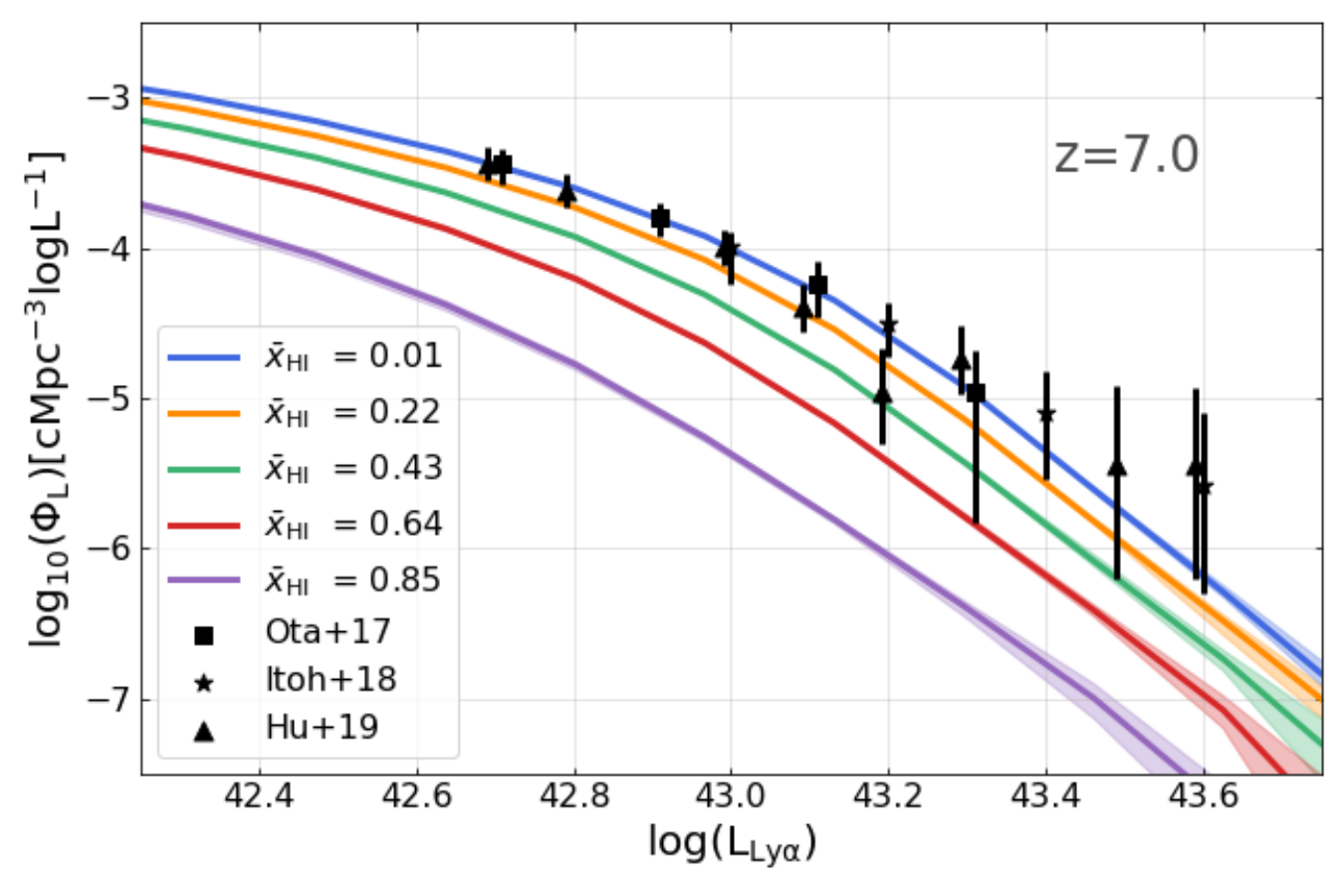}
\caption{The curves show the LAE LF after IGM attenuation for various global neutral fractions, \xf, at $z=7$. The shaded regions show the range in which 95\% of 30 simulation realizations reside. Observational data from \citet{ota2017, itoh2018, hu2019} are shown with various markers. The data have been offset by 0.01 with respect to each other on the abscissa to aid with visualization. The observational data at $z=7$ is consistent with a completely or mostly ionized IGM, as demonstrated with the data's agreement with the blue and orange curves. This, combined with the $z=5.7$ \lya\ LF agreement, shows that the evolution of the $z=5.7$ to $z=7$ \lya\ LF is explainable with smaller halos and lower intrinsic \lya\ emission, rather than an increase in global neutral fraction in our simulation. \label{fig:attenuated_z7_lfs}}
\end{figure}

Figure \ref{fig:attenuated_z7_lfs} shows the LAE LF calculated from the full simulation volume for various \xf\ values. As \xf\ increases, attenuation of intrinsic \lya\ emission increases and the resulting LAE LF is suppressed; between a fully ionized universe and a universe with \xf\ $\sim 0.85$, the suppression is about an order of magnitude. The observational data at $z=7$ agree well with a fully ionized universe, which implies that in our simulation, the evolution of the \lya\ LF from $z=5.7$ to $z=7$ in observations can be fully explained via halo growth and changing intrinsic \lya\ emission. Previous works, such as \citet{dijkstra2014a}, have argued that at least some of the \lya\ LF evolution likely comes from evolution of galaxy properties.

\section{Cosmic Variance in LAEs}\label{cosmic_variance}
\noindent

With our simulation constructed, we begin analysis to determine the importance of cosmic variance on the inference of global neutral fraction. As laid out in the introduction, \lya\ photons are easily scattered by neutral hydrogen--this has the effect of decreasing the number of observed LAEs when neutral fraction of \hi\ increases. This decreases the volume density of observed LAEs. The differential change in the \lya\ LF between redshifts has been used to infer the global neutral hydrogen fraction of the universe \citep{malhotra2004, ouchi2010, kashikawa2011, konno2014, zheng2017, ota2017, itoh2018, konno2018, hu2019, wold2021, morales2021}. Our goal is to quantify how cosmic variance impacts this inference and explore mitigation strategies in survey design with an updated, empirical LAE model.

\subsection{A Nominal LAE Survey}\label{nominal_survey}
\noindent
To gain some footing, we first consider a nominal survey, representative of volumes probed by typical surveys carried out by ground-based telescopes equipped with narrowband filters. We will ``observe'' our simulation with the specifics of the nominal survey and use Bayesian inference to estimate \xf\ from the ``observed'' \lya\ LF. 

\begin{figure}
\epsscale{1.1}
\plotone{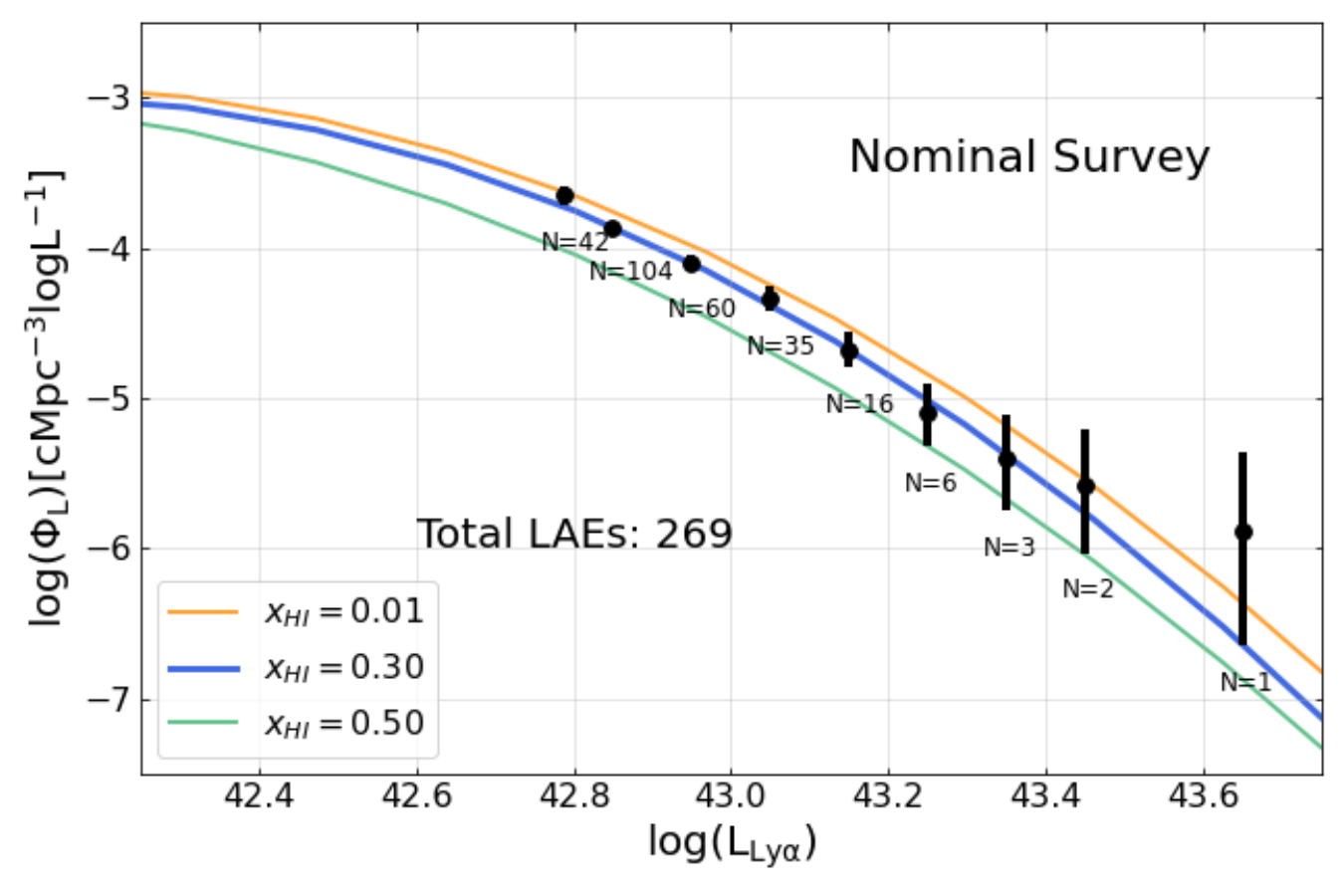}
\caption{The black points show a LAE LF measured from a 2 square degree survey with perfect observations to a depth of $f_{Ly\alpha}>1\times10^{-17}\:[\rm{ergs}\: \rm{s}^{-1} \: \rm{cm}^{-2}]$ and a redshift window of $6.75 < z < 7.25$. The model LFs for \xf\ = [0.01, 0.3, 0.5] are shown as the blue, black, and green curves, respectively. The error bars are assumed to be Poissonian for $\rm{N}>20$, otherwise they are tabulated from \citet{gehrels1986}. The intrinsic scatter of the black points about the 0.3 model LF, the input neutral fraction, comes from cosmic variance. \label{fig:nominal_lf}}
\end{figure}

We simulate a circular survey covering 2 square degrees to a \lya\ flux limit of $\rm{f}_{Ly\alpha}>1\times10^{-17}\:\rm{ergs}\:\:\: \rm{s}^{-1} \: \rm{cm}^{-2}$ (or $\rm{L}_{Ly\alpha}\gtrsim6\times10^{42} \rm{ergs}\:\rm{s}^{-1}$ at $z=7$). Such a survey pushes faint enough in flux/luminosity to detect LAEs fainter than the knee of the \lya\ LF at z=7 \citep{ota2017, itoh2018, hu2019}. This depth is comparable to those reached by narrowband surveys, and so provides an interesting baseline. Notably, we will consider a LAE ``observed'' simply if it falls into the geometric cutout and exceeds the minimum line flux limit; in this way, our survey ignores any effects of contamination or incompleteness, and so any stochasticity on the inferred values of \xf\ between surveys is the result of stochasticity in the number of observed LAEs alone. We impose a redshift window of $\Delta z\sim0.5$ centered at $z=7$ by selecting galaxies in a slice of 160 cMpc within the  simulated cube. Such a survey has a volume of $7.58\times10^{6}\:\: \rm{cMpc}^3$ at  $z=7$, a volume comparable to $z=7$ narrowband surveys which have larger areas but smaller redshift window \citep{ota2017, itoh2018, hu2019, wold2021}. 

We create a $z=7$ simulation with a realistic input global neutral fraction of \xf = 0.3 \citep{morales2021} and observe an area in a randomized location within the cube. This survey results in  a total sample of 274 LAEs. The LAE LF for one specific realization is shown in Figure \ref{fig:nominal_lf}. We split the LAEs into bins of width $\Delta \rm{log}(L_{Ly\alpha}) = 0.1$. The model \lya\ LFs (as measured from the entire simulation volume) for \xf = 0.01, 0.30 (input value), and 0.50 are shown as the blue, black, and green curves, respectively. The error bars associated with the simulated LF measurements come from Poisson statistics on the counts in each bin if $N_{\rm{LAE}} > 20$. For smaller numbers of LAEs, we use the estimates tabulated in \citet{gehrels1986} for small numer Poisson errors. 

\subsection{Inference on \xf\ from the Nominal Survey}
\label{inference}
\noindent

In order to speed up the inference on \xf, we fit a function to the model LAE LFs from Section \ref{full_lfs} as a function of \xf. More information on this analytical function can be found in Appendix \ref{appendices.model_surface}. Some example LAE LFs for different \xf\ are visualized in Figure \ref{fig:nominal_lf}. 

We use a Bayesian approach (implemented using the MCMC fitting from the \texttt{pymc3} package) to estimate the best fit hydrogen neutral fraction, \xf, given our model \lya\ LFs (Section~\ref{full_lfs}) and the simulated LF measurement (Section \ref{nominal_survey}). The 2D surface fit in Appendix \ref{appendices.model_surface} is used to create the predicted LAE volume densities, $\mu$, for a given neutral fraction. Then, the likelihood is given by

\begin{equation}
    p(\phi_{\rm{L}} \mid \bar{\rm{x}}_{\rm{HI}},\tau) = \prod_i \sqrt{\frac{\tau}{2\pi}}\rm{exp}(-\frac{\tau}{2}(\phi_{L,i}-\mu_i)^2)
\end{equation}
where the subscript $i$ denotes each of the measured luminosity bins. The unknown model parameters to be fit are the precision, $\tau$, of the observed LAE LF data and \xf. We consider a Gamma function prior on $\tau$ with $\alpha=\beta=1$ and a flat prior on \xf.

The posterior distribution function for \xf\ is then given by 
\begin{equation}
    p(\bar{\rm{x}}_{\rm{HI}} \mid \phi_{\rm{L}}, \tau) \propto \int p(\phi_{\rm{L}} \mid \bar{\rm{x}}_{\rm{HI}},\tau) p(\tau) p(\bar{\rm{x}}_{\rm{HI}})d\tau
\end{equation}
where $p(\tau)$ is the Gamma prior on $\tau$ and $p(\bar{\rm{x}}_{\rm{HI}})$ is the uniform prior on $\bar{\rm{x}}_{\rm{HI}}$.

We note that we are assuming here that the precision of all of the LF measurements, $\tau$, is the same. The results of our paper do not change if we instead assign adaptive luminosity bins to ensure that this assumption is true, that the precision between all the measurements is the same, i.e. the same number of LAEs go into each bin. We use fixed bin sizes in log($\rm{L}_{Ly\alpha}/[ergs\:\:\:s^{-1}]$), rather than these adaptive bins, for ease of comparison to measurements in literature.

\begin{figure}
\epsscale{1.1}
\plotone{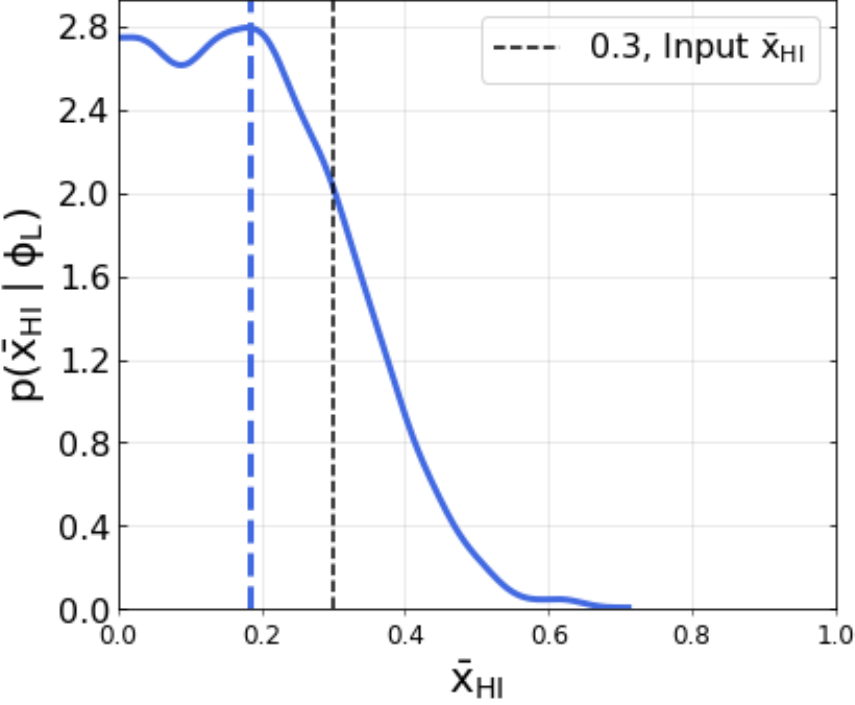}
\caption{The posterior on \xf\ inferred from the LAE LF measured in Figure \ref{fig:nominal_lf} is shown in blue. The simulation's input \xf, 0.3, is indicated with the vertical dashed black line. The posterior favors a slightly lower \xf\ than the true value because the measured region of space happened to have a slightly over-dense number of LAEs. The large width of posterior indicates that \xf\ is not very well constrained; 95\% of the posterior is contained in the range $0.01 < \bar{\rm{x}}_{HI} < 0.45$. \label{fig:nominal_posterior}}
\end{figure}

The posterior on \xf\ for the LAE LF shown in Figure \ref{fig:nominal_lf} is displayed in Figure \ref{fig:nominal_posterior}. The input neutral fraction is denoted with the vertical black dashed line. We can see that the width of the posterior is rather wide and the posterior PDF continues increasing toward low-values of \xf. The posterior's median is indicated with the vertical dashed blue line--it underestimates the real neutral hydrogen fraction. Looking back at the LAE LF in Figure \ref{fig:nominal_lf}, the origin of this offset is apparent: more of the observed LF data points lie above the input \xf\ than below it, and a higher number density of LAEs leads to an inference of a lower \xf. In other words, the global neutral fraction is underestimated because the number of LAEs in the specific realization of the nominal area is higher than the predicted number,
i.e. the randomized survey position fell on a slightly over-dense region of the simulation (with respect to LAEs).

Motivated by this result, in the following sections we explore the range of LAE densities expected at $z=7$, quantify how this range of environments propagates to a range on the inferred \xf, and then test methods to mitigate the effect of varying environments on the inference of \xf.

\subsection{The Effect of Cosmic Variance on \xf\ Inference}\label{inference_cosmic_variance}
\begin{figure*}
\epsscale{1.1}
\plotone{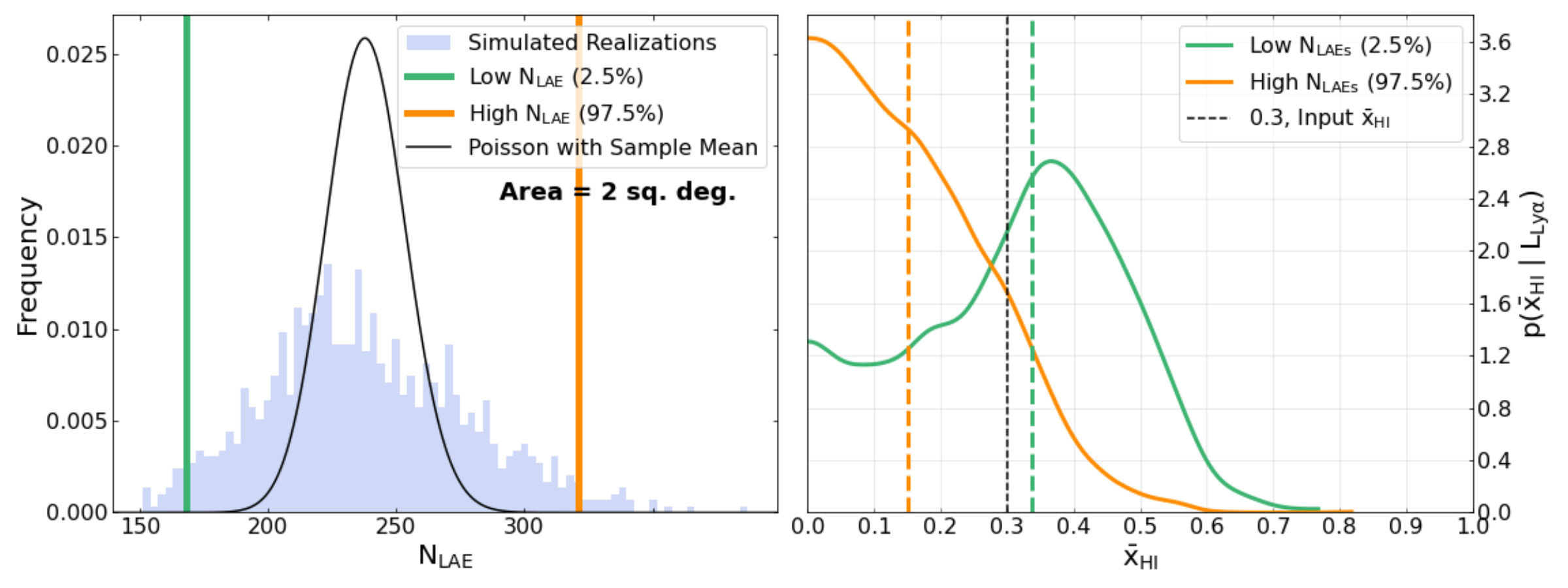}
\caption{The left panel shows the distribution of the number of LAEs observed, $N_{\rm{LAE}}$, in 10,000 realizations of the nominal 2 square degree survey in a \xf=0.3 at $z=7$ universe in blue. We investigate one over-dense region (orange vertical line) and one under-dense region (green vertical line) further. A Poisson distribution with the same sample mean is shown in black--the excess variance over the Poisson distribution is due to the clumpy nature of the large-scale-structure and the rarity of LAEs. The right panel shows the posteriors on \xf\ for the extrema of the 95th percentile range of LAE densities in orange (high LAE density) and green (low LAE density). The simulation's input \xf\ is shown with the vertical black dashed line. The difference between the medians of these distributions is $\Delta \sim 0.19$.  \label{fig:extrema_posteriors}}
\end{figure*}

\noindent
We simulate 10,000 realizations of the nominal survey scattered randomly about the simulation volume and consider the total number of LAEs, $\rm{N}_{LAE}$, in each realization which meet the detection threshold. The distribution of $\rm{N}_{LAE}$ is shown in blue in the left panel of Figure \ref{fig:extrema_posteriors}. For comparison, a Poisson distribution with the same population mean is shown in black; two particular survey realizations, one at the 2.5 and one at the 97.5 percentile of the distribution of $N_{\rm{LAE}}$, are noted with green and orange vertical lines, respectively. We will look at these particular realizations to determine a conservative (i.e. covering 95\% of surveys) absolute variance in \xf\ inferred from the different over- and under-dense environments.


We calculate the posteriors on \xf\ from the LAE LF measured from the over- and under-dense regions and show them in the right panel of Figure \ref{fig:extrema_posteriors}; orange shows the posterior on \xf\ inferred from the LAE sparse region, and green shows \xf\ inferred from the LAE rich region. The difference between the two medians is quite large, about $\Delta \Bar{x}_{\textrm{HI}} \sim 0.19$. This difference is critically important, as it is impossible to know a priori if a particular survey has landed in a region which is over-dense, under-dense, or something close to the average density, and this has a direct effect on the inferred \xf. This issue is still present for studies which use the differential evolution of the LAE LF to infer \xf, because we cannot know if the comparative surveys are in over- or under-dense regions of the universe with respect to LAEs.

This inference was done in a very ideal scenario, where there is no impurity, no incompleteness, and the model LAE LFs we are comparing to our ``observations'' are from the same simulation, and so we know we have the physics in the modelling precisely correct. In reality, there would be additional uncertainty entering from the fact that comparison models will make assumptions on the underlying physics. Despite this extremely optimistic scenario, $\Delta \Bar{x}_{\textrm{HI}} \sim 0.19$ at $z=7$ when inferring from our nominal survey's LAE LF, a rather large uncertainty. It is important to note that this uncertainty in \xf\ is a function of \xf\ itself, $\Delta\bar{\rm{x}}_{\rm{HI}}(\bar{\rm{x}}_{\rm{HI}})$. We find, in agreement with \citet{mason2018a} that the uncertainty decreases as \xf\ increases, owing to a narrower distribution of ionized bubbles sizes at high \xf\ values. We save a full quantification of $\Delta\bar{\rm{x}}_{\rm{HI}}(\bar{\rm{x}}_{\rm{HI}})$, akin to the work done for quasars and GRBs in \citet{mesinger2008b}, for future work, and instead take the uncertainty at \xf$=0.3$ as a baseline and investigate mitigation strategies.

It is noteworthy that this nominal survey is already fairly large and pushes to moderate flux depths.
Taking into account the considered redshift window, $6.75<z<7.25$, the survey's volume, $7.58\times10^{6}\:\: \rm{cMpc}^3$, is larger than many of the $z=7$ LAE surveys observed to date, and comparable to \citet{wold2021}. Larger volumes, deeper surveys, or independent fields may mitigate this systematic uncertainty, so we now explore how $\Delta \Bar{x}_{\textrm{HI}}$ changes when varying the survey strategy. 

\subsection{Mitigating the Systematics}
\noindent
We consider three scenarios to mitigate systematic uncertainty on the inference of \xf\ : increasing the survey area, increasing the survey depth, and considering a survey composed of many independent fields.
\subsubsection{Increasing Area}\label{increasing_area}
\begin{figure*}
\epsscale{1.1}
\plotone{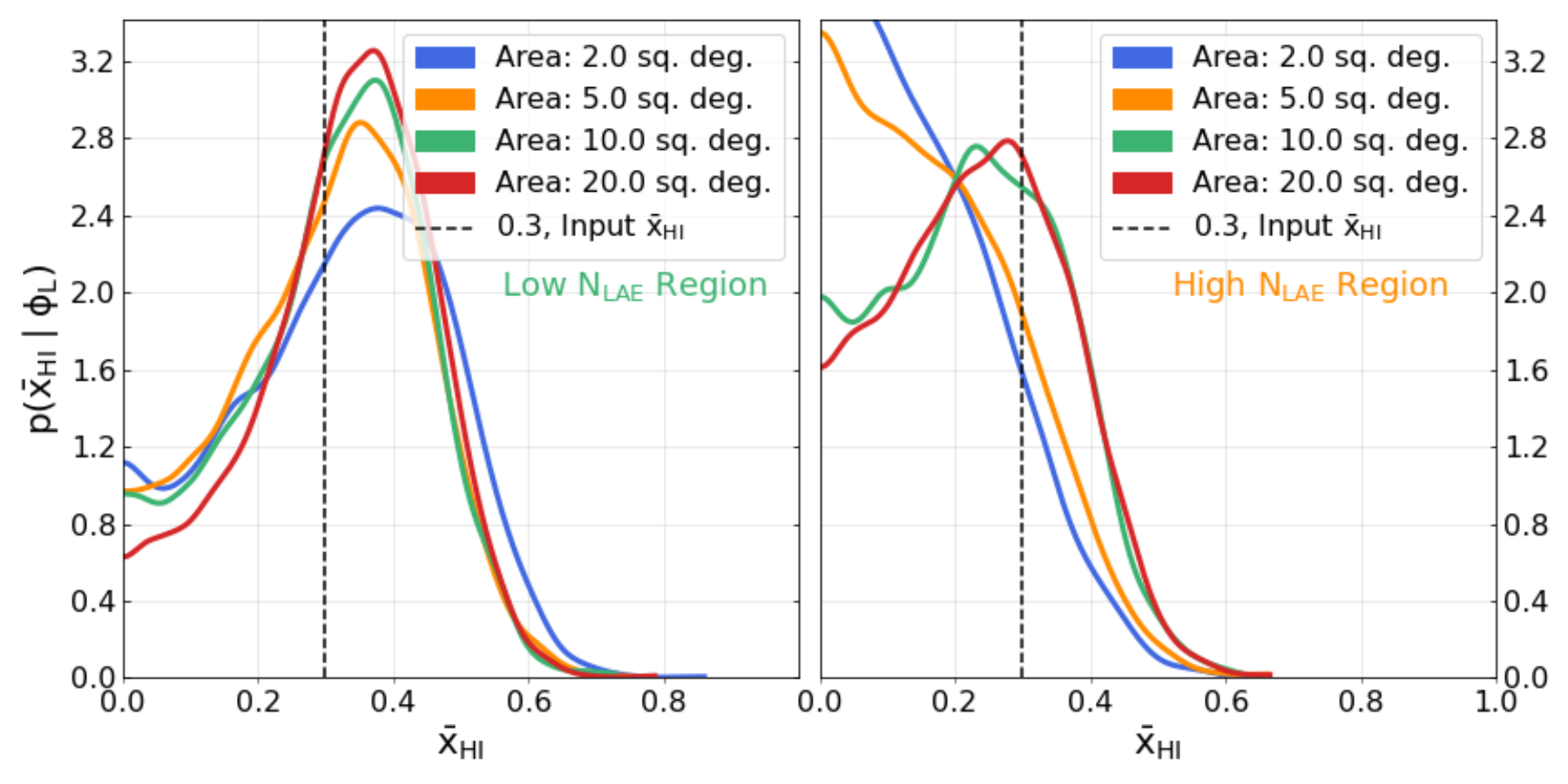}
\caption{We increase the area of the nominal surveys on the previously considered density extrema locations while keeping the luminosity limit and redshift window constant. Increasing the area to 20 square degrees centers the posterior on the true value of \xf, so we will further investigate 20 square degree surveys to determine the intrinsic spread in inferred \xf\ from cosmic variance. \label{fig:vary_area_extrema}}
\end{figure*}

\noindent
To determine survey areas which may be effective at decreasing the effect of cosmic variance on the inference of \xf, we consider four increasingly large areas centered on the two positions identified in Figure~\ref{fig:extrema_posteriors}, i.e. an over-dense and an under-dense region. 
We  simulate areas of 2, 5, 10, and 20 square degrees, while keeping the survey line-flux limit fixed to the nominal survey. For each survey, we use the same procedure described in Section \ref{inference} to infer the neutral hydrogen fraction. The resulting posteriors for both the over- and under-dense regions are shown in Figure \ref{fig:vary_area_extrema}. 

Figure \ref{fig:vary_area_extrema} shows that increasing the area by a factor of 5 to 10 has a tendency to modestly decrease the width of the posterior--in the LAE sparse region, the standard deviation of the 2 square degree survey's posterior is 0.15, while the 20 square degree survey's standard deviation is 0.13. 



\begin{figure}
\epsscale{1.1}
\plotone{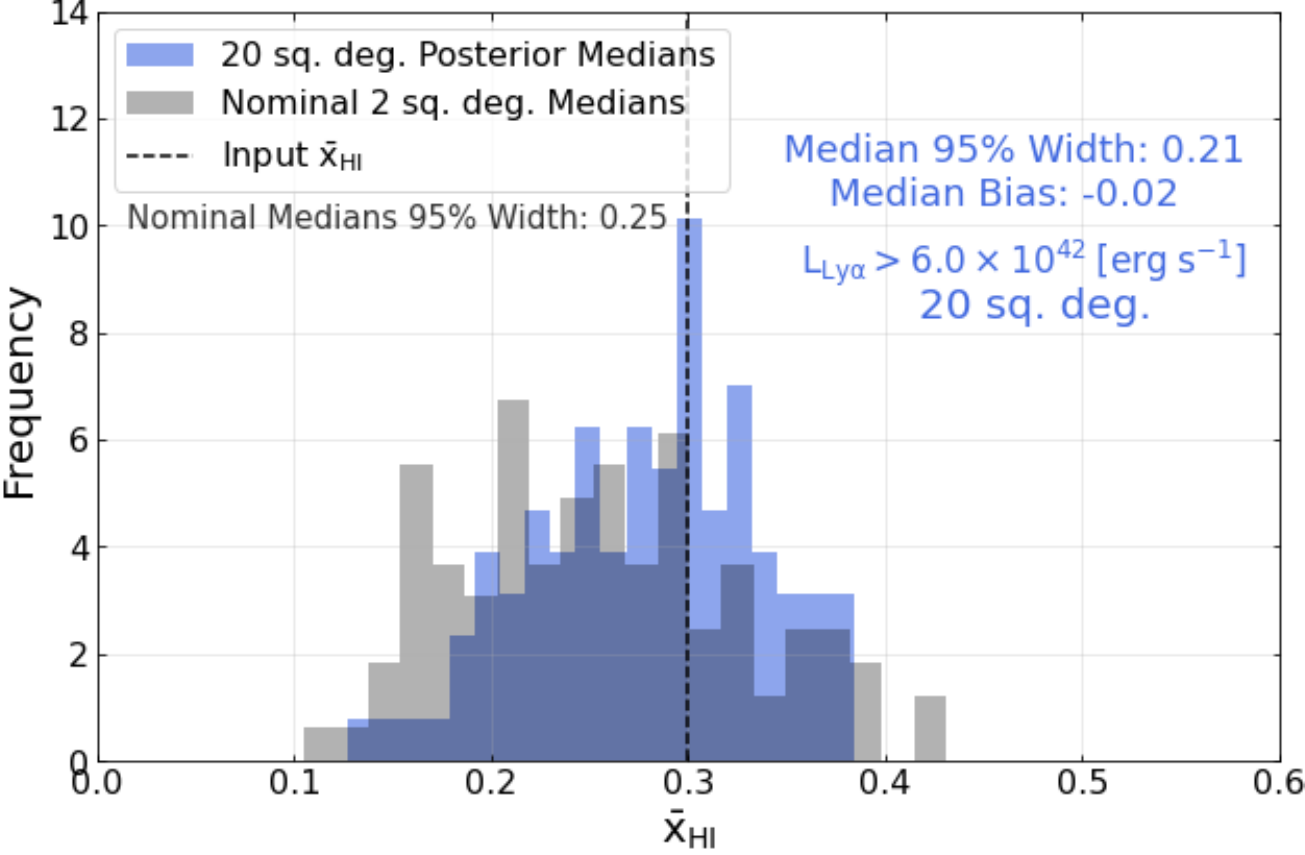}
\caption{The distribution of 100 posterior medians for the 20 square degree survey (blue) and nominal 2 square degree survey (black) for MCMC analyses. Both of these surveys have $\Delta z=0.5$ and luminosity limit $\rm{L}_{Ly\alpha} > 6\times10^{42} [\rm{ergs}\:\rm{s}^{-1}]$. The area of the 20 square degree survey was chosen because it seemed to remove the offset in inferred \xf\ with respect to the true value seen in Figure \ref{fig:vary_area_extrema}. The scatter in the posteriors medians is consistent between the two survey strategies, indicating that even increasing the area of the survey by a factor of 10 is not enough to overcome the fact that the uncertainty on \xf\ is dominated by cosmic variance and imposes an uncertainty $\Delta\bar{\rm{x}}_{\rm{HI}} \sim 0.2$. \label{fig:large_area_summary_stats}}
\end{figure}

We take the extremely optimistic case, an observationally perfect 20 square degree survey with $\Delta z=0.5$ and luminosity limit $\rm{L}_{Ly\alpha} > 6\times10^{42}\:\: \rm{ergs}\:\rm{s}^{-1}$ and investigate it further. 

We compute 100 realizations of this survey, randomly positioning them within the simulated volume, and infer \xf\ from each. We calculate the median for each of these 100 posteriors and plot the distribution in Figure \ref{fig:large_area_summary_stats}. For a baseline comparison, we repeat the same procedure with the nominal 2 square degree survey (which has the same flux limit as this 20 square degree survey); 100 realizations are created and each is used to make an inference on \xf. The medians of the 100 nominal survey posteriors are plotted in Figure \ref{fig:large_area_summary_stats} as the underlying black distribution. 95\% of the nominal survey's medians are within a range $\Delta$\xf$=0.25$, while the 20 square degree medians have 95\% of their realizations within $\Delta$\xf$=0.21$. The width of the distribution of these medians gives us a sense of the systematic uncertainty in \xf\ arising from cosmic variance between the two surveys; there is an extremely modest tightening of posteriors for the 20 square degree surveys compared to the 2 square degree surveys, which does not seem to be statistically significant.


Thus, even increasing our area by a factor of 10x to 20 square degrees, considering a fairly wide redshift window, and detecting LAEs fainter than the luminosity function knee at z=7 is not enough to significantly reduce the uncertainty on \xf. For surveys larger than 20 square degrees, we begin to get to the point where even our extremely large simulation cannot provide a large number of independent fields of view; we thus cannot conclusively say what volume is necessary to drive the uncertainty on \xf\ below $\sim0.2$ for a survey with flux limit $\rm{f}_{Ly\alpha}>1\times10^{-17}\:\rm{ergs}\:\: \rm{s}^{-1} \: \rm{cm}^{-2}$, we can only say that it is larger than $7.58\times10^{7}\:\: \rm{cMpc}^3$, a significantly larger volume than those which have been probed in LAE surveys so far. Evidently, simply taking a survey with a larger area is not enough to mitigate the effect of cosmic variance.

\subsubsection{Increasing Depth}
\begin{figure}
\epsscale{1.1}
\plotone{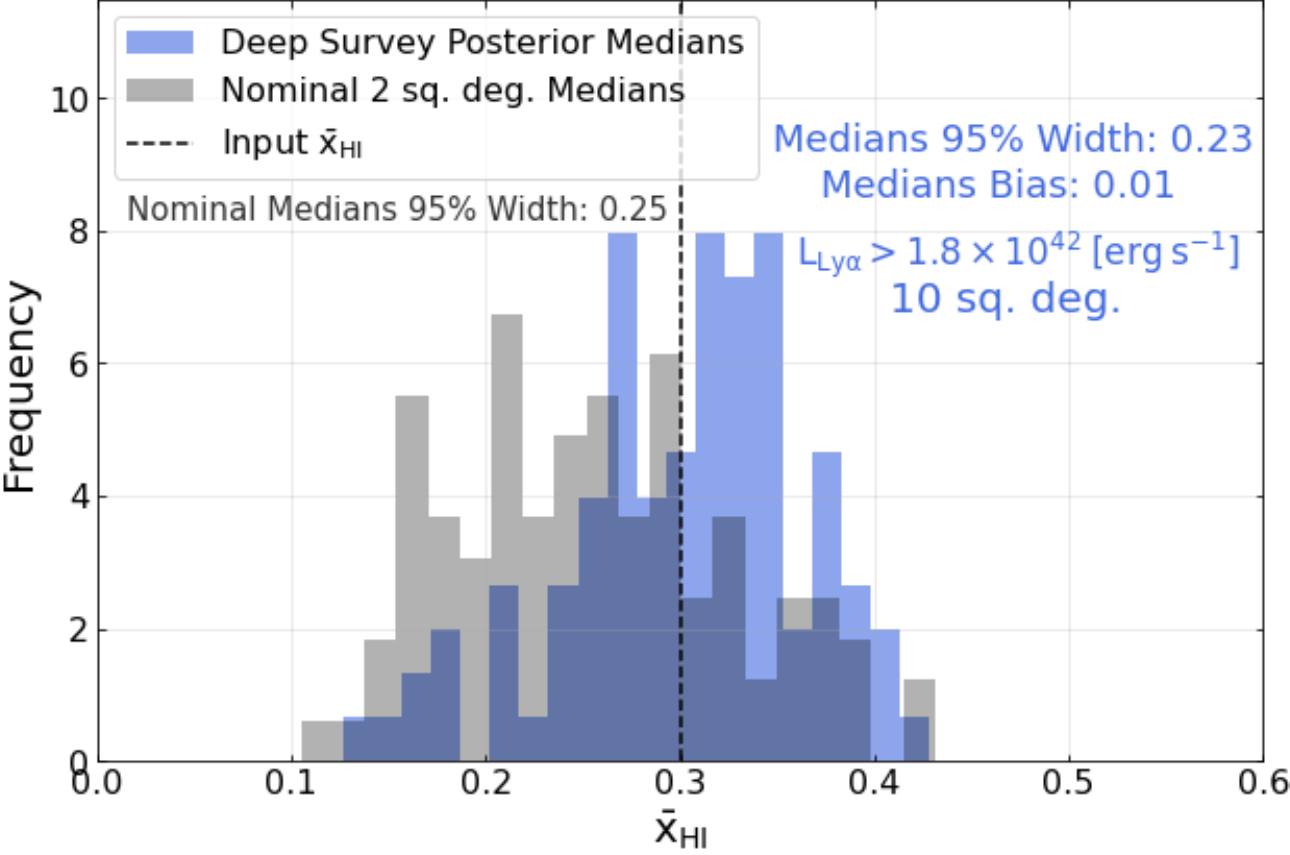}
\caption{The distribution of 100 posterior medians for the deep ($\rm{L}_{Ly\alpha} > 1.8\times10^{42} [\rm{ergs}\:\rm{s}^{-1}]$) 10 square degree surveys (blue) and the posterior medians of the 100 nominal surveys (black). For both survey strategies $\Delta z=0.5$. The widths of the distributions of both survey strategies is about the same; 95\% of the nominal survey's medians are within a range $\Delta$\xf$=0.25$, while the deep 10 square degree medians have 95\% of their realizations within $\Delta$\xf$=0.23$. This indicates that pushing to deeper flux limits is not a viable strategy in mitigating the effect of cosmic variance. \label{fig:large_deep_summary_stats}}
\end{figure}

\noindent
One may wonder what effect additional depth would have on the inference of \xf, motivated by the fact that less luminous LAEs will generally occupy less massive halos, and so be less biased, and thus be less sensitive to cosmic variance. To investigate this idea, we repeat the exercise from Section \ref{increasing_area}, but modify the survey--we instead consider a 10 square degree survey with a flux limit of $\rm{f}_{Ly\alpha}>3\times10^{-18}\:\rm{ergs}\:\: \rm{s}^{-1} \: \rm{cm}^{-2}$. This survey's flux limit is about 3.3x deeper than the nominal survey, so it would take about 11x as long to execute. Following the steps from the previous section, we calculate the posteriors on \xf\ from 100 realizations of such a survey; again, the distribution of the medians for the 100 surveys are shown in Figure \ref{fig:large_deep_summary_stats}. Again, the distribution for the 100 nominal surveys is shown in black as a baseline to compare against.

The width which encompass 95\% of the posterior medians for the deep survey is virtually the same as the nominal survey, $\Delta$\xf$=0.23$. Thus, even pushing $\sim3.3$x deeper in flux and increasing the area by 5x over the nominal survey, requiring $\sim$55x more exposure time, has not decreased the systematic uncertainty.


\subsubsection{Independent Fields}\label{ind_fields}
\noindent
Lastly, we consider a survey composed of many independent fields, the motivation being that each field will probe a different environment for the LAEs, averaging out high- and low-density environments. 

\begin{figure*}
\epsscale{1.1}
\plotone{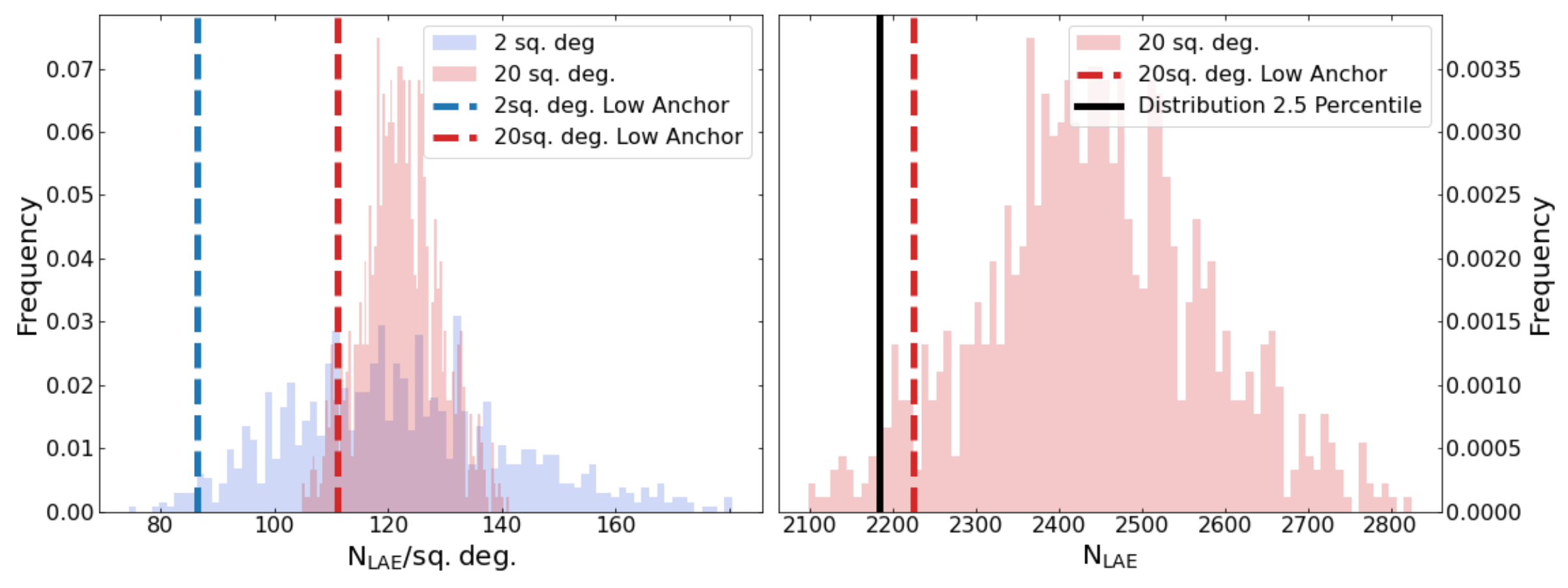}
\caption{The left panel shows the distribution of the number of LAEs per square degree observed by 1000 surveys of areas 2 and 20 square degrees (and to depth $\rm{L}_{Ly\alpha}\gtrsim6\times10^{42} \rm{ergs}\:\rm{s}^{-1}$) in blue and red, respectively. Surveys for each area, centered on the under-dense region identified in the nominal 2 square degree survey, are indicated with the dashed vertical lines of the same colors. The distributions of $\rm{N}_{LAE}$ per square degree tend to narrow as survey area increases. The right panel shows the distribution of the $\rm{N}_{LAE}$ observed in 1000 realizations of the 20 square degree. The dashed red vertical line again indicates the under-dense region selected with the 2 square degree survey, and the distribution's 2.5 percentile is indicated with the solid black vertical line. The imprint of the under-dense region identified with a 2 square degree survey is still dominating the 20 square degree survey's total number of LAEs compared to other 20 square degree surveys, illustrated by the fact that the red dashed line does not significantly move towards the distribution's mean. \label{fig:added_area_analysis}}
\end{figure*}

Further motivation can be seen examining Figure \ref{fig:added_area_analysis}. The left panel shows the distributions of the number of LAEs per square degree observed by 1000 surveys of areas 2 and 20 square degrees (and to depth $\rm{L}_{Ly\alpha}\gtrsim6\times10^{42} \rm{ergs}\:\:\rm{s}^{-1}$). The blue dashed line indicates a 2 square degree survey realization centered on an under-dense region (at the 2.5 percentile) and the red dashed line shows a 20 square degree survey \textit{centered on that same position}. As expected, the distributions of the number of LAEs per square degree converges towards the overall simulation mean as the survey area increases; the square degree surveys' distribution (red) is far narrower than the two square degree surveys' (blue).

However, the imprint of the under-dense region identified with a 2 square degree survey is still dominating the number of LAEs observed in the 20 square degree survey when compared to the parent distribution, as shown in the right panel. The 20 square degree survey centered on the under-dense region (dashed red vertical line) is extremely close to the parent distribution's 2.5 percentile (solid black vertical line). 

This is indicative of the fact that adding area to a given survey has a tendency to add the average number of LAEs per square degree, by definition. This will have no effect on moving a given realization towards the center of the distribution of $\rm{N}_{LAE}$ for the parent distribution of that survey size; only adding an under-dense or over-dense region changes its position relative to the parent distribution. Then, a survey centered on an under-dense (over-dense) region will continue to exhibit low (high) LAE counts relative to other similar surveys, even as the area increases drastically.

Instead, we simulate a survey of 100 independent 0.2 square degree fields, totalling 20 square degrees, to a depth of $\rm{f}_{Ly\alpha}>3\times10^{-18}\:\rm{ergs}\:\: \rm{s}^{-1} \: \rm{cm}^{-2}$, the same total area as our largest survey and as deep as our deepest simulated survey. Each 0.2 square degree field has the inference done on it separately, and the resulting 100 posteriors are multiplied together to form a joint posterior on \xf.

\begin{figure}
\epsscale{1.1}
\plotone{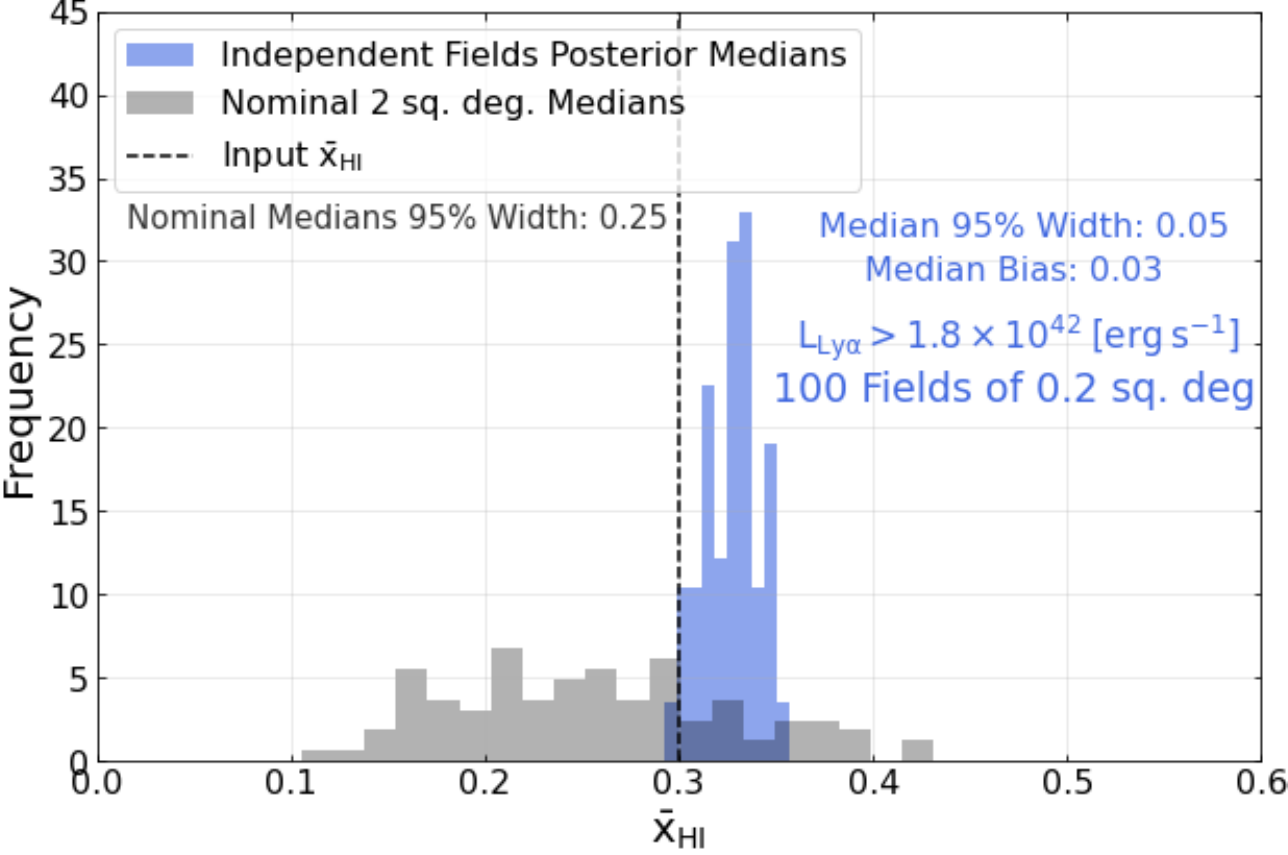}
\caption{The distribution of posterior medians for 90 surveys composed of 100 0.2 square degree fields to a depth of $\rm{L}_{Ly\alpha}\gtrsim 1.8\times10^{42} \rm{ergs}\:\rm{s}^{-1}$ is shown in blue. The distribution of the 100 nominal survey's medians is shown in black. 95\% of the nominal survey's medians are within a range $\Delta$\xf$=0.25$, while the 90 surveys composed of 100 independent fields have 95\% of their realizations within $\Delta$\xf$=0.05$. This strategy is effective in reducing the systematic uncertainty on \xf. \label{fig:mega_passage}}
\end{figure}

Running a survey in this particular manner is computationally expensive, and so we limit ourselves to 90 realizations of such a survey; we show the distribution of the posterior medians in Figure \ref{fig:mega_passage}. The width which encompasses 95\% of the surveys has decreased--$\Delta$\xf$=0.05$. Probing many independent fields is the only effective way to decrease the systematic uncertainty introduced by cosmic variance, decreasing the 95\% width by about a factor of 4.

This method has an additional advantage; \citet{mesinger2008} showed that counts-in-cell statistics can be an effective way to use the enhanced clustering of LAEs during reionization to constrain \xf. While the counts-in-cells method does not strictly require that the fields be discontiguous, a survey strategy such as ours has the additional benefit of overcoming any effect of cosmic variance, which could impact a clustering analysis as well. Thus, such a survey would have two methods of constraining \xf, both of which are robust against cosmic variance, which could be checked against each other for consistency. 

\subsubsection{The Implication of the Mitigation Techniques}
\noindent
We are left to conclude that the only effective way to break through the floor in the uncertainty in \xf\ arising from the cosmic variance of LAEs is with surveys composed of many independent fields. Drastically increasing the survey area and pushing to much deeper fluxes alone proved ineffective. Of note, the survey composed of independent fields, with a flux limit of $\rm{f}_{Ly\alpha}>3\times10^{-18}\:\rm{ergs}\:\: \rm{s}^{-1} \: \rm{cm}^{-2}$ and an area of 20 square degrees would take $\sim110$ times as long to carry out as the nominal survey, which is currently representative of the largest dedicated LAE surveys to date--thus, this serves more as a proof of concept than a realistic goal for a survey design in the immediate future.

It is worth keeping in mind that an uncertainty of $\Delta$\xf$=0.2$, as we have demonstrated is characteristic when the universe has \xf$=0.3$, is still a competitive constraint. Particularly, if such measurements are made at a range of redshifts, we may be able to construct a fairly constraining timeline of reionization; thus, the LAE LF will continue to be a useful tool in constraining reionization, in addition to other techniques, such as damping wings in quasars and GRBs, LAE clustering, LAE fraction, and \lya\ EW distributions.


\section{Conclusions}\label{conclusion}
\noindent
We have post-processed a $z=7$ comoving $1.6\textrm{Gpc}^3$ simulation to create a mock LAE catalog using empirical relations. Our simulation simultaneously reproduces the $z=6$ and $z=7$ UV LFs and the $z=5.7$ \lya\ LF.

We produce $z=7$ \lya\ LF models, ranging from $42 < \rm{log(L_{Ly\alpha}/[ergs\:s^{-1}]) < 44}$, as a function of \xf. We created mock surveys with different observation strategies and compare the measured \lya\ LFs to these models, inferring posteriors on \xf. The observed LAE LF varies, deriving from the large scale structure and stochasticity in the amount of neutral gas along the sightline between the LAEs and the observer, and the shape and position of the peak posterior distribution function on \xf\ changes in turn. 

We show that the precision of \xf\ estimates drawn from an LAE LF derived from a single 20 square degree field with $\rm{f}_{Ly\alpha}>1\times10^{-17}\:\rm{ergs}\:\: \rm{s}^{-1} \: \rm{cm}^{-2}$ is limited by the cosmic variance of LAEs. We find an uncertainty floor of $\Delta\bar{\rm{x}}_{\rm{HI}} \sim 0.2$ when \xf\ $=0.3$. Note that a 20 square degree survey with a redshift window of $\Delta z=0.5$ has a volume of $7.58\times10^{7}\:\: \rm{cMpc}^3$, about 10 times larger than existing narrowband LAE surveys.

We investigated three methods to push this floor in the uncertainty down. 
(1) Increasing the area covered by a factor of 10, (2) pushing to $\sim3.3$x deeper flux limits, and (3) breaking up the survey into independent fields. For the added area and deeper survey strategies, the variance in the medians of the posteriors on \xf\ remains virtually unchanged from the nominal 2 square degree survey. This demonstrates that the \xf\ inferred from observed LFs in contiguous fields are largely at the whim of what cosmic environment one's survey happened to land in.

However, a large, deep survey composed of smaller independent fields proved effective in reducing the systematic uncertainty on \xf. Under this strategy, we see width of 95\% of the posterior medians of \xf\ decrease from $\Delta\bar{\rm{x}}_{\rm{HI}} \sim 0.2$ to $\Delta\bar{\rm{x}}_{\rm{HI}} \sim 0.05$. Thus, probing multiple independent fields is critically important in constraining \xf. 

This result demonstrates that surveys of LAEs aiming at constraining \xf\ should adopt a strategy similar to the one proposed in Section \ref{ind_fields} in order to minimize the systematic uncertainty on \xf. The approved JWST survey Parallel Application of Slitless Spectroscopy to Analyze Galaxy Evolution (PASSAGE, PI M. Malkan), as a pure parallel survey, follow this strategy. It will be carried out during Cycle 1 of JWST observations and its observations of high-z LAES will allow the inference of new information about the timeline of reionization.

\section{acknowledgements}
The authors would like to thank Masami Ouchi and Kazuaki Ota for sharing their \lya\ LF observations. Also, a thanks to Steven L Finkelstein for sharing his conglomerated UV LF data, a very useful resource. Lastly, a very special thanks to Micaela Bagley, who provided useful discussions at key points throughout this project.
SB and CS were partially supported by a Jet Propulsion Lab/NASA grant to the University of Minnesota (RSA 1677370 / NNN12AA01C). M.H. is supported by the Knut and Alice Wallenberg Foundation. CM acknowledges support by NASA Headquarters through the NASA Hubble Fellowship grant HST-HF2-51413.001-A awarded by the Space Telescope Science Institute, which is operated by the Association of Universities for Research in Astronomy, Inc., for NASA, under contract NAS5-26555, and support by the VILLUM FONDEN under grant 37459. The Cosmic Dawn Center (DAWN) is funded by the Danish National Research Foundation under grant DNRF140.
\newline

\noindent
\textit{Software:} \texttt{Astropy} \citep{robitaille2013}, \texttt{SciPy} \citep{oliphant2007}, \texttt{NumPy} \citep{vanderwalt2011}, \texttt{Matplotlib} \citep{hunter2007}, \texttt{emcee} \citep{foreman-mackey2013}, \texttt{pymc3} \citep{salvatier2016}, and \texttt{arviz} \citep{kumar2019}.

\clearpage
\appendix\section{Luminosity Function Analytical Fit}\label{appendices.model_surface}
We assume the following 2D functional form, which reproduces the logs of the modelled \lya\ LFs:

\begin{equation}
\begin{aligned}
    \rm{log_{10}}(\Phi) = \beta_0 + \beta_1 x + \beta_2 y + \beta_3 x^2 + \beta_4 x^2y + \beta_5 x^2y^2\\ + \beta_6 y^2 + \beta_7 xy^2 + \beta_8 xy   
\end{aligned}
\end{equation}

\noindent 
where $x$ is the log of the \lya\ luminosity and $y$ is \xf. The luminosity functions are in units [$\rm{cMpc}^{-3}\:\rm{logL}^{-1}$] and the luminosities are in units [$\rm{ergs}\:s^{-1}$]. The $\beta$ coefficients are reported below and are unitless. The fit is valid in the range $0.01 < \bar{\rm{x}}_{HI} < 0.91$ and $ 42 < \rm{log}\:(L_{Ly\alpha}/[\rm{ergs}\:s^{-1}]) < 44$.

\begin{table*}[ht]
\caption {Beta Parameters} 
    \centering
    \begin{tabular}{|c|c|c|c|c|c|c|c|c|}
     \hline
    $\beta_0$ & $\beta_1$ & $\beta_2$ & $\beta_3$ & $\beta_4$ & $\beta_5$ & $\beta_6$ & $\beta_7$ & $\beta_8$ \\
     -2436.6103 & 115.69888 & -2563.8187 & -1.3750958 & -1.4037818 & 3.340239 & 6146.6548 & -286.62831 & 119.99238 \\ 
    \hline
    \end{tabular}
    \label{tab:fit_parameters}
\end{table*}

\clearpage

\bibliographystyle{aasjournal} 
\bibliography{My_Library2}

\end{document}